\documentclass[]{aa}
\usepackage{graphicx}
\usepackage{natbib}
\usepackage{journals}

\def \be {\begin{equation}}
\def \en {\end{equation}}
\def \bea {\begin{eqnarray}}
\def \ena {\end{eqnarray}}

\def \bi{\begin{itemize}}
\def \ei{\end{itemize}}

\def \eg {{\it e.g. }}
\def \ie {{\it i.e. }}

\begin{document}

%***********************************************************************************
\title{Competition between shocks and entropy floor : unifying groups
and clusters of galaxies} 
%***********************************************************************************

\author{S. Dos Santos\inst{1} \and O. Dor\'e\inst{2}}
\offprints{S. Dos Santos}
\institute{Department of Physics \& Astronomy, University of
Leicester, 1 University Road, Leicester LE1 7RH, UK\\
\email{ssa@star.le.ac.uk} \and  Institut d'Astrophysique de Paris, 98
bis boulevard Arago, F-75014 Paris, FRANCE\\
\email{dore@iap.fr}}
\date{}
%
%========================================================================

\abstract{Semi-analytic models of X-ray clusters and groups of galaxies,
relying on the idea that there was a non-gravitational energy injection in
these systems, are able to reproduce many observed correlations, in
particular the $L_X-T$ relation and the ``central entropy floor'' in
groups. Limiting models exist which describe the behaviour of clusters and
groups separately, but no analytic modeling has yet been found to unify both
mass ranges. {\it It is the aim of this paper to provide such an analytic
model.} Our description relies on a now standard description of the shock
thought to occur in these systems near the virial radius \citep{CMT98}, the
isothermality and spherical symmetry of the intracluster medium, as well as
the reinterpretation of observed quantities (like the X-ray luminosity, the
gas mass $M_{ICM}$ and the central SZ effect $y_0$) in terms of the specific
entropy. This allows the derivation of analytic expressions for several
observed correlations ($L_X-T$, $M_{ICM}-T$, $y_0-T$,...) and their
normalisation encompassing \emph{both the group and the cluster regimes}. The
analytic predictions compare very well with observations, as well as with
more elaborated semi-analytic schemes.  This agreement allows a
reinterpretation of the $L_X-T$ relation (via the quantity $L_X/T^{7/2}$) and
the $y_0-T$ relation (via $y_0/T^{5/2}$) as indirect measures of the
non-gravitational entropy content of groups and clusters of
galaxies. Finally, we emphasize the need for shock heating, even in the
group mass range : \emph{shocks can not be completely
supressed in groups} (and thus groups can not be entirely isentropic)  unless
an unacceptably high entropy floor is needed in 
order to break the self-similarity in the $L_X-T$ relation.
}

%fg-t
%m-t

\titlerunning{Unifying groups and clusters with entropy}
\authorrunning{S. Dos Santos \& O. Dor\'e \qquad } 
\maketitle
%
%========================================================================

%========================================================================
\section{Introduction}
%========================================================================

For ten years, it has been known that X-ray observations of clusters of
galaxies can not be reproduced in simple self-similar models, where the
central gas density is proportional to dark matter density. \citet{Kaiser91}
and \citet{EH91} advocated a preheating of the gas before it fell into the
cluster potential to recover the observed correlations. Later, X-ray
observations of groups of galaxies have strengthened the case for a
non-gravitational entropy injection in these systems, in particular by the
observation of the so-called ``entropy floor'' \citep{PCN99}. These authors
showed that the central entropy level in groups (outside the cooling-flow
radius) is higher than the level the sole cosmological shocks can provide,
while being well in accord with adiabatic numerical simulations ({\it
e.g.\/,} including no dissipative physics) in clusters. This entropy
injection could be due to supernovae explosions (the so-called feedback from
star formation) or other sources as active galactic nuclei. Several
semi-analytic schemes that elaborate on this idea have been proposed to
reproduce the curvature of the $L_X-T$ relation in the groups' mass range
\citep[where the effects of the preheating are thought to be the highest,
see][]{CMT97,Bower97,VS99,ValSchaef2000,WFN2000,BBLBCF2000}. In particular,
\citet[][hereafter CMT]{CMT97,CMT98,CMT99} have introduced a simple model,
where the bending is provided by the differential strength of a shock
occuring at the virial radius in a preheated gas. Their description relies on
the physical modelling of the shock interface, using Hugoniot relations, and
reproduces naturally the central observed entropy floor. While this model
highlights the important role of shocks and entropy floor in the formation of
clusters and groups, no {\it analytic} description of the X-ray observed
relations has yet been found to encompass both groups and clusters.
\emph{The aim of this paper is to present such a model and compare it to
observations.}

However, despite the success of these semi-analytic schemes when confronted
to observations, difficult problems remain, as for example the fact that a
reheating by supernovae (hereafter SN) explosions requires an incredibly high
efficiency of the transfer of energy from SN remnants to the intergalactic
medium \citep{VS99,BBLBCF2000}. In particular, the cooling of the SN remnants
will obviously decrease this transfer efficiency, and must be ignored in
these models. While a combination of SN and quasars (hereafter QSOs)
reheating is probable and would alleviate this problem, \citet{Bryan2000} has
proposed that differential galaxy formation between groups and clusters
(which would lower the central entropy in groups and allow higher entropy gas
to flow into the center) can explain the curvature of the $L_X-T$ relation
and the entropy floor. However, it is well known that without preheating,
most of the available gas in the universe would have formed stars today,
which is not observed \citep[this the so-called ``overcooling problem'',
see][]{BVM92,BPBK2001}. This much-needed feedback in galaxy formation would
certainly have an impact on the formation of clusters and
groups. \citet{VS99} have indeed shown that galaxy and cluster feedback
differing requirements are likely to provide a tight constraint on the amount
of preheating in the universe. But, even in the preheated scenarios, there is
actually no consensus on the entropy injection epoch. While most of the
studies have focussed on ``external preheating'' models (where the entropy
injection occurs before the formation of groups and clusters {\it e.g.\/,}
before the gas is compressed by shocks) because low density gas requires much
less energy than high density one to be raised to a given entropy level (and
also because the star formation rate seems to peak rather early in the
universe evolution), \citet{Loewenstein2000}, from a series of approximate
static hot gas models and \citet{BM2001}, from 1D numerical simulation with
cooling, mass dropout and star formation feedback, have argued that most of
the heating occured during or after the assembly of the group or cluster
gas. The efficiency problem of SN explosion is still present, but can be
alleviated by an initial mass function flatter than the Salpeter one
\citep{BM2001}. These are the so-called ``internal preheating'' models.

In this paper, we focus explicitly on an external preheating model. Internal
preheating models require spatially-dependent and time-dependant star
formation rates to provide the amount of injected entropy, while external
preheating models only require the level of entropy before the gas falls in a
cluster or a group. This simplification allows us to obtain a completely
analytic model that describes both groups and clusters.The physical approach
we follow consists of deriving scaling relations, \eg $L_X-T$, $y-T$, $y-L_X$
by linking these observed quantities to the specific entropy profile in the
system.  For this purpose we first derive an \emph{analytic expression for
the entropic profile normalisation at the virial radius} of clusters and
groups by discussing shocks, as in CMT. Using this relation we then derive
analytically the relevant scaling relations and compare them to data.
Following \citet{TN2001}, we highlight the key role of entropy and argue that
it indeed constitutes the best observable in clusters, allowing to derive
analytical expressions for standard X-ray and Sunyaev-Zeldovich (hereafter
SZ) correlations. This leads us to reinterpret these relations in terms of
global entropy content in a well physically motivated manner.

The plan of the paper is as follows : in section~\ref{sec:shocks}, we present
the shock model and derive an expression for the normalisation of the entropy
profile at the virial radius. This expression, containing a free
normalisation, is fitted to observations. In section~\ref{sec:lxt}, we relate
the X-ray luminosity to the entropy profile and derive an analytic expression
for the shape of the $L_X-T$ relation, which is subsequently compared to
observations \emph{using only the fitted parameter of the last section}. The
same method is followed in sections~\ref{sec:fgas}, 
\ref{sec:yt} and \ref{sec:ylx} to obtain analytic expressions for the
$M_{ICM}-T$, $f_{gas}-T$, $y_0-T$ and $y_0-L_x$ correlations (where $M_{ICM}$
is the gas mass, $f_{gas}$ is the gas fraction and  $y_0$ is the central
Compton parameter). In each of these parts, we provide a detailed comparison
with previous work. Section~\ref{sec:disc} discusses a reinterpretation
of these correlations in terms of the total entropy content in groups and
clusters, tests it using a semi-analytic model and discusses the main
hypotheses we make. Moreover, a comparison of our results for the entropy
floor with other theoretical models is made. Section~\ref{sec:conc}
summarizes the present work and concludes. We also compute the numerical
values of the normalisations of the correlations found and summarize all
these results in Appendix~\ref{app:a}, allowing to use them easily in another
context. Appendix~\ref{app:b} computes the infalling velocity as a function
of the mean mass of the system and compares to hydrodynamic
simulations. Finally, Appendix~\ref{app:shapefactor}, introducing a simple
model for the entropy profile of groups, computes the shape
factors appearing in the different normalisations.

Unless otherwise stated, we use $H_0 = 100 \, h_{2/3} \, \rm km \, \, s^{-1}
\, \, Mpc^{-1}$ with $h_{2/3} = 2/3$, which gives approximatly $H_0 = 67 \,
\rm km \,\, s^{-1} \,\, Mpc^{-1}$. We rescale all the data used in this paper
to this value.

%========================================================================
\section{Entropy and shocks in clusters and groups}
%========================================================================
\label{sec:shocks}

The study of the equilibrium state of X-ray clusters and groups of galaxies
requires the physical description of the complex interplay between baryonic
and non-baryonic dark matter. To tackle this problem analytically requires
several simplifying hypotheses. The 
following assumptions are reasonable and will be thus assumed henceforth
\citep[see][for a review and a discussion of the main assumptions]{Sarazin88}. 
% The complexity of the interplay between the DM and the ICM
% cannot be yet described in its full glory. Nevertheless the following reasonnable
% assumptions will be assumed henceforth (see Sarazin 1988 for a
% review). 

First, the hot and diluted plasma of ions and electrons constituting the
intracluster medium (hereafter ICM) is treated as a single species perfect
gas satisfying  
% First the ICM is regarded as a hot and dilute plasma of ions and electrons
%  (considered as a single fluid) whose low density and high temperature allows
%  to be treated as a perfect gas satisfying thus
\be
\label{eq:eqstat}
P = \frac{\rho kT}{\mu \, m_p} \: ,
\en
where $P$, $\rho$, $T$ denote respectively the pressure, density and
temperature of the baryonic gas, while $k$ denotes the Boltzman constant.
Second, neglecting the ICM mass with regards to the DM mass, and assuming
{\it stationarity} (no gravitional potential variation on time scales smaller
than the hydrodynamic time scale, \eg no recent mergers), the ICM is assumed
to be in {\it hydrostatic equilibrium} (hereafter HSE) in the dark matter
gravitational potential $\phi$. We can thus write the continuity and the
Euler equation as follows~:
\bea
\nabla(\rho v) & = & 0 \label{eq:continuity}\\
\nabla P & = & -\rho \nabla \phi \label{eq:eqhydro}
\ena 
where $v$ is the gas velocity and $\phi$ the gravitational
potential. Equations~(\ref{eq:continuity}), (\ref{eq:eqhydro}) and
(\ref{eq:eqstat}), together with the energy conservation equation and
Poisson equation (the use of which can be avoided if an approximate analytic
expression for $\phi$ is chosen, {\it e.g.\/,}  neglecting the gravitational
contribution of the baryons) form a closed set of equations, whose
resolution provides the radial profiles of all the quantities of interest in
the gas.  

To solve this set, boundary conditions are required. In the present paper, we
are interested in the variations of these boundary conditions with the total mass
of the system (or equivalently with its mean temperature). Indeed, the
previous equations will describe the equilibrium of baryonic gas in the
underlying potential, whatever the total mass of the system. We argue and
show in the following that changes in the boundary conditions, when explicitly
written as a function of the mean temperature, allow us to derive the change
in global correlations, such as the $L_X-T$ relation, on a scale going from
clusters to groups of galaxies. For this purpose, we use the fact that the
flow of gas on a cluster becomes supersonic in the vicinity of the virial
radius $r_v$ \citep{Te97}. We will then show that a   
standard modeling of the {\it resulting shock} (following CMT) leads to
the appropriate boundary conditions. We will consider hereafter that the
boundary conditions take place after the last major merger of the system, as
appropriate in a hierarchical picture of structure formation.

The last hypothesis we will use is that within the virial radius, the
ICM is {\it isothermal}. This hypothesis will not be used
in the derivation of the boundary conditions (sec.~\ref{sec:shockgen}) but
in the derivation of the global correlations in groups and clusters
(sec.s~\ref{sec:lxt} and~\ref{sec:yt}). Both observations \citet{MFSV98}
and simulations \citep{Te01} show that this approximation is good into  
a factor two and is thus sufficient for our purpose. 

Finally, let us recall an hypothesis necessary underlying every work published
using the entropy of the ICM, but rarely stated : {\it the local
thermodynamic equilibrium hypothesis} (hereafter LTE). By
definition, in a state of global thermodynamic equilibrium,
temperature and pressure (and thus density) are constant throughout
the system, and the state functions of the system (\eg entropy) have a
determined dependency on the state variables chosen (and so a fixed
value). In the LTE however, temperature and pressure can vary from point to
point (as is obvious in clusters from X-ray observations for example) and,
locally only, the system is in thermodynamical equilibrium. Thus, the state
functions have the same dependency on state variables as in global
equilibrium, but vary now from point to point as do the latter. Note
that this hypothesis is likely to break down at smaller scales than the
ones of interest to us, due to turbulence or magnetic effects. This
hypothesis is crucial if we want to use  
the usual analytic expression of the {\it specific entropy} of the ICM, 
$s$, namely  
\be 
s = S/c_v = \ln \left(\frac{k T}{\rho^{\gamma -1}}\right) 
\en 
where $\gamma$ is the polytropic index of the gas.
 The thermodynamical state of the gas is fully known
as soon as the entropy profile of the gas is known. Since we want to
highlight the key role of entropy as ``an observable'' we choose to
describe the physical state of clusters in terms of entropy, or more
appropriately and equivalently in terms of the \emph{adiabat}, defined as
\be 
\label{eq:kdef}
K = \left(\mu \, m_p\right)^{\gamma-1} e^s = \frac{k T}{n^{\gamma -1}} =
\left(k T\right) n^{-2 / 3}\ ,   
\en
where $n = \rho/(\mu \, m_p)$. In the following, $T$ will denote the
temperature in keV ({\it i.e.\/,} we will replace everywhere $k T$ by $T$). 
Note that from now on the polytropic $\gamma$ index will be fixed to its
standard value of $5/3$. The definition of the adiabat differs from the
widely used definition 
of \citet{BBP99} (by a constant factor $(\mu \, m_p)^{\gamma-1}$) but
allows a direct comparison to observations. 

\subsection{A general shock model}
%*********************************
\label{sec:shockgen}

We first aim at deriving adiabat boundary conditions based on the
Rankine-Hugoniot relations, used in the same form as CMT.

\subsubsection{The Rankine-Hugoniot relations}
% --------------------------------------------

Let the infalling gas velocity reach the sound speed at $r=r_v$ and
consider subsequently that a shock form at this radius. Let the
temperature, the density and the velocity of the infalling gas before
the shock be respectively $T_1$, $n_1$ and $v_1$, and the post-shock
temperature, density and velocity be $T_2$, $n_2$ and $v_2$. If the
shock efficiency is very high, \ie the post-shock velocity
$v_2$ is null \emph{in the rest-frame of the cluster center of mass},
the Rankine-Hugoniot relations yield (CMT)~: 
\be
\label{eq:posttemp}
k T_2 = \frac{\mu \, m_p \, v_1^2}{3} \, F(\epsilon),
\en
with
\be
\label{eq:fepsilon}
F(\epsilon) = \frac{(1+\sqrt{1+\epsilon})^2}{4} + \frac{7}{10}\, \epsilon -
\frac{3}{20}\frac{\epsilon^2}{(1+\sqrt{1+\epsilon})^2}  
\en
and 
\be
\label{eq:epsilondef}
\epsilon = \frac{15}{4}\frac{kT_1}{\mu \, m_p \, v_1^2} \; .
\en
At the same time, the ratio between post-shock ($n_2$) and pre-shock ($n_1$)
density is given by : 
\be
\label{eq:postdens}
\frac{n_2}{n_1} = 2 \left(1-\frac{T_1}{T_2}\right) + \left[ 4
\left(1-\frac{T_1}{T_2}\right)^2 + \frac{T_1}{T_2} \right]^{1/2}\: .
\en
Note that since the ICM within $r_v$ is assumed to be isothermal, its
temperature is equal to the temperature $T_2$ after the
shock. 

Using the formulas above, we will now derive a general expression for the
adiabat jump at the shock, which will depend on the system mean
temperature. This virial adiabat normalisation will provide us with the
change in the boundary conditions necessary to describe groups and clusters
in a unique analytic model.

\subsection{The adiabat after the shock}
%---------------------------------------

\begin{figure} 
   \centering 
   \includegraphics[width=9cm]{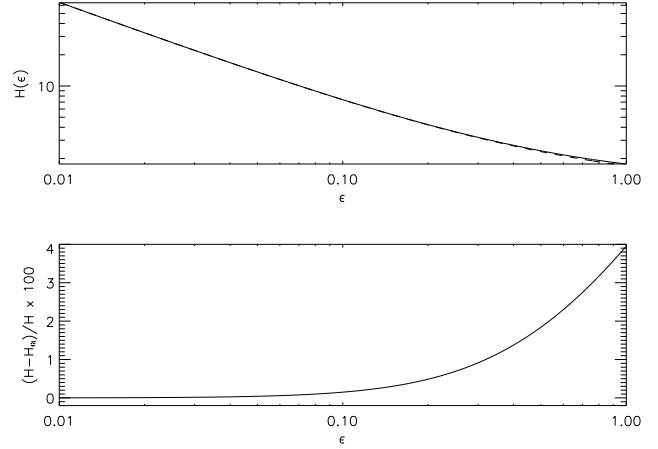}
   \caption{Top : plot of the function $H(\epsilon)$
   (equation~\ref{eq:hepsilon}, solid line), together with the two
   firts order of its expansion (equation~\ref{eq:fith}, dashed line). Both
   functions are hardly distinct. Bottom : Percentage difference
   between $H$ and its expansion. The agreement is better than $4\,
   \%$ in the relevant $\epsilon$ range (and always better than $15\, \%$ for
   all values of $\epsilon$.)}
   \label{fig:happrox} 
   \end{figure}

To find a general expression for the post-shock adiabat, we
first express the ratio $T_1/T_2$ as a function of $\epsilon$ :
\be
\frac{T_1}{T_2} = \frac{4}{5}\frac{\epsilon}{F(\epsilon)}.
\en
Introducing this expression in equation~(\ref{eq:postdens}), one has :
\bea
\label{eq:rhoepsilon}
\textstyle n_2 & = & 2\, n_1 \, \Bigl(
\bigl(1-\frac{4}{5}\frac{\epsilon}{F(\epsilon)}\bigr) \nonumber \\ 
& + & \qquad \left[ \left( 1-\frac{4}{5}\frac{\epsilon}{F(\epsilon)}\right)^2
+ \frac{\epsilon}{5 \, F(\epsilon)}\right]^{1/2} \Bigr). 
\ena
Using then equations~(\ref{eq:posttemp}), (\ref{eq:epsilondef}) and
(\ref{eq:rhoepsilon}), the post-shock adiabat $K_2$ is easily expressed
as a function of the pre-shock adiabat $K_1$ and $\epsilon$ in the following
manner~: 
\bea
K_2 & = & \frac{k \, T_2}{n_2^{2/3}} \nonumber \\
  & = & \frac{5}{2^{8/3}} \, K_1 \, H(\epsilon),
\ena
where 
\bea
\label{eq:hepsilon}
\textstyle H(\epsilon) & = & \frac{F(\epsilon)}{\epsilon} \, \Bigl(
\bigl(1-\frac{4}{5}\frac{\epsilon}{F(\epsilon)}\bigr) \nonumber \\ 
& + & \qquad \left[ \left( 1-\frac{4}{5}\frac{\epsilon}{F(\epsilon)}\right)^2
+ \frac{\epsilon} {5 \, F(\epsilon)}\right]^{1/2} \Bigr)^{-2/3}\: . 
\ena   
Even if this last expression looks complicated, it can be
straightforwardly expanded in a Laurent serie in the vicinity of $\epsilon =0$. We
obtain~:
\bea
\textstyle H(\epsilon) = {17 \over 10}\ 2^{-2/3} \left( 1 + {10 \over
17}\ {1\over \epsilon} \right) + \mathcal{O}(\epsilon)\ .
\label{eq:fith}
\ena
Checking the validy of this expansion in the range
$0~\le~\epsilon~\le~1$ we find an agreement better than $4\%$
(figure~\ref{fig:happrox}) which is enough for our purpose (the error of the
approximation reaches a constant $15\, \%$ level  at large
$\epsilon$). Thus we will neglect higher order corrections.  

%-----------------------------------------------------------
%-----------------------------------------------------------

As a consequence, for the all range from groups to rich clusters, one can write to a very
good approximation~:
\be
K_2 =  \frac{17}{16} 2^{-1/3}\, K_1 \, \left( 1 + {10 \over 17}{1 \over
\epsilon}\right)\: .
\label{eq:kdvt}
\en
Note that this expression that fits so well the general adiabat expression
is not really a surprise. Its physical significance is indeed
straightforward. To show this, let us take two limiting behaviour, namely in
the cluster and in the group mass range.

Since rich clusters of galaxies accrete mainly small clumps and diffuse gas,
the \emph{cold inflow} or \emph{strong shock} limit is
appropriate. Consequently it is usually argued that the entropy of
rich clusters is dominated by shock created entropy. Indeed,
for this approximation we have $T_2 \gg T_1$ and $\epsilon
\ll 1$, and so (from  eq.~\ref{eq:kdvt})
\be
K_2 \simeq \frac{5}{8}{2^{-1/3} \over \epsilon}\, K_1\: .
\en

On the other hand, for groups of galaxies, the \emph{weak shock} or
\emph{adiabatic infall} approximation is more appropriate. Indeed
since the infall speed tends to be lower, and since groups accrete
preheated  gas, the weak shock limit (whose limit is the true adiabatic
infall) is appropriate (CMT)\footnote{Note that this hypothesis of adiabatic
infall has been shown to allow to reproduce the $L_X-T$ relation in the group
range by \citet{BBP99}}. Indeed, taking the limit $\epsilon \rightarrow 1$, 
eq.~(\ref{eq:kdvt}) yields 
\be
K_2 \simeq {27 \over 16}2^{-1/3}\, K_1\: .
\en
This means that the entropy content of groups is
dominated by the ``adiabatically accreted'' gas, hence the
existence of a so-called entropy floor (the last expression of $K_2$ being
independent of $\epsilon$).   

The analytic expression in eq.~(\ref{eq:kdvt}) thus provides  a well
physically motivated expression for the competition between shocks and
entropy floor that  rules the adiabat virial normalisation from groups to
clusters.  
   
\subsection{Linking groups and clusters}
%***************************************

\subsubsection{Defining a general analytic $K-T$ relation}
%---------------------------------------------------------

We now want to make use of the previously derived
expression~(\ref{eq:kdvt}) to predict a general relation for the $K-T$
relation for groups and clusters. 

First, making use of eq.~(\ref{eq:epsilondef}) and eq.~(\ref{eq:kdvt})  we rewrite
\be
K_2 =  \frac{17}{16}2^{-1/3}\, K_1 \, \left( 1 + {8 \over 51}{\mu \, m_p \,
v_1^2 \over k \, T_1}\right)\: .
\en

Second, we need to express the infall velocity $v_1$ as a function of the
mean temperature of the system. This is done in Appendix~\ref{app:a},
where we show that using the virial theorem, the $M-T$ relation,
together with the assumption of adiabatic cold inflow, we can write
($\phi_1$ being a caracteristic virial radius gravitational potential,
$T_*$ a caracteristic temperature and $\eta$ a dimensionless constant, see
eq.~\ref{eq:v1}, \ref{eq:mt} and \ref{eq:phivt}) : 
\be
\label{eq:v1t}
v_1^2  = -2 \eta \phi_1 \left(\frac{T}{T_*}\right)  \;. 
\en
Note that from now on we will assume notations of
Appendix~\ref{app:a}.

This leads naturally to:
\be
\label{eq:k2tcomplete}
K_2(T) =  \frac{17}{10}2^{-2/3}\, K_1 \, \left[ 1 - {16 \over 51} \eta {\mu \, m_p \,
\phi_1 \over k \, T_1} \left(\frac{T}{T_*}\right) \right]
\en
that we rewrite
\be
\label{eq:knorm}
K_2(T) = K_0 \left[ 1 + \left( \frac{T}{T_0} \right)\right]\: ,
\en
where  we have defined\footnote{Remark that $17/16 \times 2^{-1/3} \simeq 0.843$, so that
$K_0 \simeq 0.843 \, K_1$.}
\be
\label{eq:k0}
K_0  =  \frac{17}{16}2^{-1/3}\, K_1 
\en
and
\be
\label{eq:t0}
T_0 = - \frac{51}{16 \, \eta} \frac{k \, T_1}{\mu m_p \, \phi_1} T_* \: .
\en

\emph{The scaling relation~(\ref{eq:knorm}) is a key result of our paper}. It states how the
entropy jumps due to the shock changes with the mean temperature of the
system. Written this way, the last formula has two free parameters~: a normalisation
parameter $K_0 \simeq K_1$, \ie the adiabat before the shock, and the
temperature $T_0$ that marks the transition from the adiabatic inflow to the
strong shock regime. 

Deriving the value of $K_1$ requires a whole semi-analytic
scheme, which would require a model of the entropy evolution of the
intergalactic medium (hereafter IGM) as the universe evolves
\citep[see {\it e.g.\/}][]{VS99}. This is clearly beyond the scope of
the present paper, and we will thus constrain $K_1$ by comparison with
observations (which only means that the normalisation of our model is
left free). However, $K_0$ and $T_0$ are not independent parameters and
can be related by simple physical considerations as will be shown in the next section. 

Once this is done, the scaling relation~(\ref{eq:knorm}) is left with \emph{only
one normalisation parameter}~$K_0$.

\subsubsection{Relating $K_0$ and $T_0$}
%--------------------------------------------
\label{sec:compt0}

First, to explicit the relation between $K_0$ and $T_0$, we need to
specify the value of both $\eta$ and $T_1$ in
equation~(\ref{eq:t0}). Using the results of \citet{Miniati2000} for
the infall velocity in hydrodynamic simulations (see eq.~\ref{eq:vsim}
and Appendix for notations), a value for $\eta$ can be derived by equating
eq.~(\ref{eq:vsim}) and eq.~(\ref{eq:v1t})~:   
\be
\label{eq:eta}
\eta = - \frac{v_{s0}^2}{2 \,\phi_1} \left(\frac{T_*}{T_s}\right)
\en   
which, when introduced in equation~(\ref{eq:t0}), gives~:
\be
\label{eq:t0_2}
T_0 = \frac{51}{8} \frac{k \, T_1}{\mu m_p \, v_{s0}^2} \, T_s\; ,
\en
where $T_s$ and $v_{s0}$ are known (appendix~\ref{app:a}).

The relation between $K_0$ and $T_0$ will appear when expliciting
$T_1$. Indeed we can write it as a function of the pre-shock adiabat
$K_1$~:
\be
\label{eq:t1}
T_1 = K_1 \, n_1^{2/3}\; .  
\en 
The overdensity that eventually collapses and reaches a density $\rho_1$ at
$r_v$ has decoupled from the Hubble expansion at a turn-around redshift
$z_{ta}$ where its density was $\rho_{ta}$ and its radius was
$r_{ta}$. Following \citet{BBP99}, we assume first that the IGM was preheated
(non-gravitationally) before $z_{ta}$ and its adiabat raised to the value
$K_1$. Second, we also assume that between $r_{ta}$ and the shock radius
$r_v$ itself, the gas has an isentropic behaviour.  To fix the value of the
overdensity (and thus the density) $\delta_{ta} =
(\rho_{ta}-\bar{\rho})/\bar{\rho} \displaystyle$ at the turn-around, we will
consider the simplest collapse model, \ie the spherical collapse model in a
given cosmology.

%The simplest non-linear model of the evolution of a density perturbation
%(which will eventually form a group or a cluster) is the
%spherical model, where the overdensity decouples from the Hubble expansion at
%turn-around ($z = z_{ta}$, where the radius of the perturbation is $r_{ta}$)
%and begins to collapse. Let us follow  \citet{BBP99} in assuming that
%the IGM was preheated before $z_{ta}$, its adiabat being raised at the
%value $K_1$. Between $r_{ta}$ and the shock radius $r_v$ itself, the
%gas is assumed to have an isentropic behaviour. 
By definition of the baryonic fraction $f_B$ and the overdensity
$\delta_{ta}$ we can write~:
\be
\label{eq:rho1}
\rho_{ta} = (1+\delta_{ta}) \, \rho_c \, f_B 
\en 
where $\rho_c$ is the critical density. Inserting this last
equation in eq.~(\ref{eq:t0_2}), one obtains~:    
\be
\label{eq:t0fin}
T_0 = \frac{51}{8} \frac{K_1 \, \left((1+\delta_{ta}) \, \rho_c \,
f_B\right)^{2/3}}{\left(\mu m_p\right)^{5/3} v_{s0}^2} \, T_s\; ,
\en
which relates $K_0$ and $T_0$ via eq.~(\ref{eq:k0}).  

Consequently, to fix the numerical value of $T_0$ we need to specify a
value of $K_1$. We choose to assume the value of the
observed entropy floor visible in groups of galaxies. A value of
$K_1 = 120 \, \rm keV \, cm^2$ is inferred from the
data \citep[see the next section]{LPC2000}. 

Then we assume a flat $\Omega_m = 1$ universe with $H_0 = 50 \, \rm
km \, s^{-1} \, Mpc^{-1}$ and the baryonic fraction as fixed by Big-Bang
Nucleosynthesis standard, \ie $f_B = 0.015 \, h_{100}^{-2}/\Omega_m$
\citep{OSW2000}. 
For a spherical collapse model $\delta_{ta} = 9\pi^2/16$. Note however that
in the widely used hierarchical structure formation paradigm, this monolithic
formation model is too 
simplistic. Indeed, since the  majority of the infall of group of
galaxies occurs through filaments, an overdensity of $\delta_{ta} \sim
10$ should be more appropriate.  Considering then the values of
$v_{s0}$ and $T_s$ specified in the Appendix, we find
that $T_0 = 1.86 \, \rm keV$ for $\delta_{ta} = 9\pi^2/16$ and $T_0 =
2.62 \, \rm keV$ for $\delta_{ta} = 10$. This is quite
insensitive to the adopted cosmology, since for an $\Omega_m = 0.3$ universe,
one finds $T_0 = 1.71 \, \rm keV$ and $2.33 \, \rm keV$ respectively. It 
is thus obvious that we can take for $T_0$ an intermediate value
between these two extremes.   

In the following, we will thus consider that \emph{$T_0$ is fixed at 2 keV,
and that the only free parameter of the model will be the normalisation
factor $K_0$.} Moreover, choosing a value of $T_0$ between 1.5 and 2.5
keV does not alter the quality of the fits we present in the next sections.     

%Of course, $T_0$ depends in fact on the value of $K_1$ (and thus on
%$K_0$) and injecting eq.~(\ref{eq:t0fin}) into eq.~(\ref{eq:knorm})
%allows to have a expression of the post-shock adiabat wich depends
%only on the pre-shock observed value, which can then be fitted to
%observations (see the next section). 

Looking at eq.~(\ref{eq:knorm}), $T_0$ has a simple physical
interpretation : it is the system mean temperature where the effect of 
an entropy floor begins to change the adiabat jump during the virial
shock. For systems with $T < T_0$, the pre-shock entropy is high enough to
reduce the shock and a quasi-adiabatic inflow takes places (but the shocks are
not completely suppressed, see section~\ref{sec:compmodels}). For $T>T_0$, the
shock adiabat jump brings the post-shock adiabat to values much higher that
the entropy floor, the latter having less and less incidence as T increases. 
Thus we find that this transition occurs around $2\ \mathrm{keV}$.

\subsubsection{Comparing with observations}
%******************************************
\label{sec:ktobscomp}

X-ray observations of groups and clusters of galaxies have brought a great
amount of information on the gas equilibrium in these systems. Unfortunately,
due mostly to the high background of the instruments and the gas density
decline with radius, there are not yet entropy observations available up to
the virial radius, that would allow to test directly our model (and more
generally the scheme invented by CMT). The new
generation of satellites (Chandra and XMM-Newton) is expected to bring new
observations near the virial radius, allowing to test different formation
scenarios of clusters and groups \citep[see {\it e.g.\/,}][]{TSN2000}.

%-----------------------------------------------------------
   \begin{figure} 
   \centering 
   \includegraphics[width=9cm]{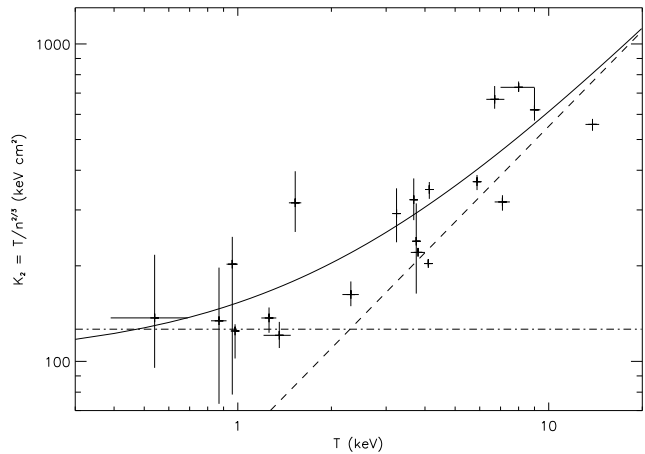}
   \caption{Post-shock adiabat $K_2$ as a function of the mean temperature of
   the system. The solid line is obtained with equation~(\ref{eq:knorm}),
   with $K_1 = 120 \rm \, keV \, cm^2$ ({\it i.e.\/,} $K_0 = 102 \rm \, keV
   \, cm^2$). Note that only the normalisation is a
   free parameter, as the reference temperature $T_0 = 2 \, \rm keV$ is
   obtained from the value of $K_1$. The dashed line is a $K \propto T$
   fit to the 
   systems above $4 \, \rm keV$ found by \citet{HP2000}. The similarity of
   the shock model 
   adiabat normalisation (computed at the virial radius) with the data taken
   at the center ($R = 0.1 \, r_v$) allows us to postulate an approximate
   self-similarity in the specific entropy profile in the Sec.~\ref{sec:lxt}.}
   \label{fig:ktrelation} 
   \end{figure}
%-----------------------------------------------------------

Nevertheless, some data on the central adiabat value of the gas in groups and
clusters are available. \citet{PCN99} were the first to show that groups have
a central adiabat higher than the one they would have if it had been imposed
by shocks only. Recently, \citet{LPC2000} have refined these observations,
taking into account the temperature gradients, while increasing the
statistics. This is the so-called observed ``entropy floor''. While cool
systems have a constant central adiabat, hotter clusters follow naturally the
results of numerical simulations without a preheating phase (which predict
that $K \propto T$). Note that our model predicts exactly
the same behaviour, exhibited in the post-shock adiabat analytic expression of
eq.~(\ref{eq:knorm}). However, we compute the adiabat near the virial radius,
while the data come from the central parts of the systems. Nevertheless,
observations and theory can be compared if we assume that the adiabat
difference between two different temperature systems \emph{is conserved when
going from the virial radius to the central part}. This implies that the
physical mechanism responsible for the post-shock adiabat value (competition
between shocks and an entropy floor) is the same as the one responsible for
the central adiabat value. This is indeed a reasonnable assumption, since 3D
hydrodynamical simulations of the formation of clusters show that, after the
last major merger, a quasi-spherical shock forms and expands, that leaves
behind an increasing adiabat profile \citep[see {\it
e.g.\/,}][]{Evrard90,Frenketal99}. The same physics is thus at work in the center
and in the outer parts of clusters, allowing the physical description of the
two regions to coincide. Indeed, the only difference between these two
regions is that the universe will have evolved when the shock reaches
the present-day virial radius, as compared to the time when the central
entropy is set. We thus don't expect the observational value of $K_1$ (found
by normalisation to the observations) to be
representative of the present adiabat of the universe, but it should instead be
representative of the IGM at the time when the smaller mass systems
formed. If virial radius observations of the adiabat were available for the
same systems, we expect then that the same analytic expression as in
eq.({\ref{eq:knorm}) would be a good description of the data, with a
different value of $K_1$, representative of the present-day adiabat of the IGM. 

Figure~\ref{fig:ktrelation} shows the comparison of eq.~(\ref{eq:knorm}) with
data from \citet{LPC2000}. Only the normalisation was fitted here, with a
value of $K_1 = 120 \, \rm keV \, cm^2$ (giving $K_0 = 102 \, \rm keV \,
cm^2$), while we have taken $T_0 = 2 \, \rm keV$, derived with the same value
of $K_1$ in the last section. Replacing eq.~(\ref{eq:t0fin}) in
eq.~(\ref{eq:knorm}) instead, and fitting
for $K_1$ gives an undistinguishible analytic curve. The dashed line is
the result of a $K \propto T$ fit to the systems above $4 \, \rm keV$ found
by \citet{HP2000}. This temperature dependence of the adiabat is
taken from numerical simulations \citep[see][]{PCN99}. The dot-dashed line is a
constant gas entropy fitted to the four lowest temperature systems by
\citet{LPC2000} and has a value of $139 \, \rm keV \, cm^2$. The agreement between
the analytic formula and the data is remarkable. Indeed, using the
fitted value of $K_1$, the computed $T_0$ value agrees very well with the
temperature of the intersection between these two limiting behaviours (see
last section). This shows that eq.~(\ref{eq:knorm}) is not a mere analytic
formula (which could have been infered from the observations), but succeeds
in capturing both ingredients which fix the central
entropy value : the entropy floor (dot-dashed line in the groups mass range)
and the shocks (dashed line, in the cluster mass range).

By adjusting the normalisation of our analytic model using the data we derive
a value for $K_1$. Then making use of this relation we get directly for any
system at a given temperature $T$ (either a group or a cluster) the
normalisation of the adiabat profile. It is now important to check wether we
are able to use this relation to derive realistic $L_X-T$ and $y-T$ relations
for both groups and clusters.

\section{From the adiabat to the $L_X-T$ relation}
%**************************************************
\label{sec:lxt}

The total X-ray luminosity of a local cluster (group) of galaxies is expressed as~:
\be
L_X = \int_{\rm V} n_e^2 \Lambda_e(\Delta E,T_e) \, \rm{dV},
\label{eq:lx}
\en
where $n_e$ is the electronic density and the integration is carried out over
the whole volume V of the system. $\Lambda_e(\Delta E,T_e)$ denotes the local
X-ray emissivity of the cluster gas within a given energy band $\Delta
E$. For example, in the case of pure thermal Brehmsstrahlung emission observed with
contemporary wide-band satellite, one has $\Lambda_e \propto T_e^{1/2}$.

\subsection{An analytic prediction for the $L_X-T$ relation}
%-----------------------------------------------------------
\label{sec:anallxt}

We can now express equation~(\ref{eq:lx}) as a function of $K$ and thus link
the X-ray luminosity with the temperature and the adiabat~: 
\be 
L_X = \int_V \frac{T^3}{K^3} \, \Lambda_e(\Delta E,T) \, 
\rm dV\: .
\en 
Since we assume the cluster to be isothermal and since in general the emissivity can be
written as a power-law function of the temperature with index $\alpha$, one
has : 
\be 
\label{eq:intk3}
\frac{L_X}{T^{3+\alpha}} =  \Lambda_0 \,\int_V
\frac{\rm dV}{K^3},
\en
$\Lambda_0$ being the normalisation of the emissivity.\\

Equation~(\ref{eq:intk3}) gives us a direct link between $L_X$, $T$ and
$K$. Was the expression of $K$ as a function of $T$ known, it would
enable us to predict the shape of the $L_X-T$ relation. However,
the result of the previous section (\ref{eq:knorm}) gives us the change in the adiabat
\emph{normalisation} as a function of temperature, but since the
integral in eq.~(\ref{eq:intk3}) is computed over all the system
volume, in principle we need the adiabat profile to compute it. However, if we 
make the further assumption that {\it the adiabat profiles are self-similar in
temperature}, \ie the temperature enters the analytic expression of
$K({\bf r})$ by its normalisation only, then we can derive an expected $L_X-T$
relation. This assumption is justified by the fact that our
theoretical expression for the adiabat normalisation
(equation~(\ref{eq:knorm}), which is computed at the virial radius, provides
a very good fit to the central entropy data in groups and clusters. This
simple fact ensures that the difference in normalisation between two
clusters of different temperature (which is due to shocks and preheating) is
leaved approximately unchanged from the virial radius to the center. This
important side aspect of our work will be discussed in
Sec.~\ref{sec:disc}. We thus write~:  
\be
\label{eq:kprof}
K({\bf r},T_2) = K_2(T_2) \, f({\bf r}),
\en
where $K_2(T_2)$ is given by equation~(\ref{eq:knorm}) and $f({\bf r})$
is a function of the radius, that is independent of the mean system
temperature. We have thus :
\be
\label{eq:lxt3}
\frac{L_X}{T^{3+\alpha}} = \Lambda_0
\,K_2(T)^{-3}\,\int_V \frac{\rm dV}{f^3({\bf r})}.  
\en 
%-----------------------------------------------------------
   \begin{figure} 
   \centering 
   \includegraphics[width=9cm]{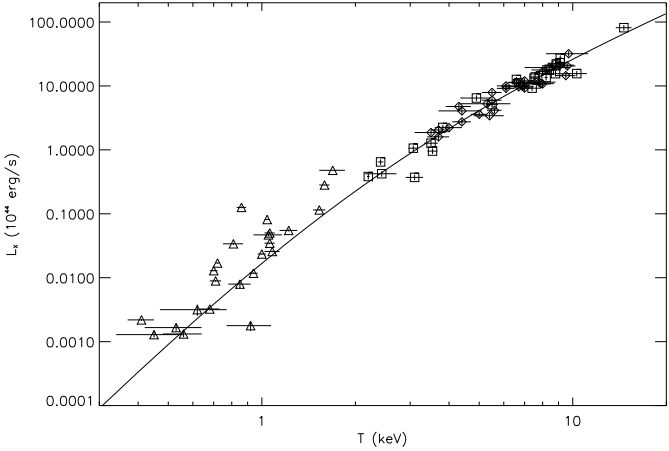}
   \caption{$L_X-T$ relation versus observations from \citet{HP2000} for
   groups and  \citet{Markevitch98} and \citet{AE1999} for
   clusters. The solid line is the analytic formula in eq.~(\ref{eq:lxt}).
   $T_0$ was fixed at the value computed in 
   Sec.~\ref{sec:compt0} {\it i.e.\/,} $T_0 = 2 \, \rm keV$, and $Q_X =
   0.115$, as computed in Appendix~\ref{app:shapefactor}.}  
   \label{fig:lxtrelation}
   \end{figure}
%-----------------------------------------------------------

Even if $f$ does not depend on $T$, the integral on the right-hand
side of the last equation depends on the total volume of the cluster,
and thus on its virial radius, which depends also on the mean
temperature. Let us assume that the cluster is spherically symmetric (which
is reasonable if the system has sufficient time to relax). We can then
write~:  
\bea
\int_V \frac{\rm dV}{f^3({\bf r})} & = & 4 \pi \int_0^{r_v} r^2
f^{-3}(r) dr \\ 
& = & 4 \pi r_v^3 \int_0^1 x^2 f^{-3}(x) dx \\
& \equiv & 4 \pi r_v^3\ Q_X
\ena 
where we set $x=r/r_v$ and define the shape factor $Q_X$. Using then
the $r_v-M$ and $M-T$ relation of equation~(\ref{eq:rvm}) and
(\ref{eq:mt}) before using our $K-T$ relation (\ref{eq:knorm}), we obtain when replacing in
equation~(\ref{eq:lxt3})~: 
\be
\label{eq:lxt}
L_X  =  L_0 \, T^{9/2+\alpha}
\left[ 1 + \left(\frac{T}{T_0}\right) \right]^{-3}
% L_X \propto T^{9/2+\alpha} \left[ 1 +
%  \left(\frac{T}{T_0}\right) \right]^{-3}
\en
where the normalisation $L_0$ is given by~:
\be
\label{eq:lxnorm}
L_0 = \Lambda_0 \left(\frac{3 \, M_*}{\Delta \, \rho_c^0 \, T_*^{3/2}} \right)
K_0^{-3} \, Q_X, \; .
\en
Let us first comment on the shape of the $L_X-T$ relation.
While for the group range, where $\alpha = 0$ and $T \ll T_0$, we have :
\be
L_X \propto T^{9/2} \propto T^{4.5},
\en
for very hot clusters (for
which $T \gg T_0$ and $\alpha = 1/2$), we recover the so-called
self-similar behaviour : 
\be
L_X \propto T^{9/2 + 1/2 - 3} \propto T^{2}\: .
\en
The 10 keV cluster range will be intermediate between these two behaviours. 
We now turn to a comparison with observations.

%  This agreement allows us
% to predict the normalisation of this 
% correlation, using only the observed value of $K_0$, the other
% parameters, $T_0$ and $Q_X$, being computed independently.

\subsection{Comparing with $L_X-T$ observations}
%-----------------------------------------------
\label{sec:lxtobs}

The normalisation of the local $L_X-T$ relation (eq.~\ref{eq:lxnorm}) depends
on the value of the IGM adiabat $K_0$ and the shape factor $Q_X$. The former
has been obtained in Sec.~\ref{sec:ktobscomp} ($K_0 = 102 \, \rm keV \,
cm^2$) by comparing our analytic formula for the adiabat jump to
observations. The latter is more difficult to obtain since it will depend on
the exact adiabat profile of clusters and groups. We show in
Appendix~\ref{app:shapefactor} that we can compute a value for $Q_X$
by considering aa reasonable model for the adiabat profile, relying on
isothermality and a $\beta-$ model for the gas density profile
\citep{CFF76}. The computed value is $Q_X = 0.115$. 

To validate this relation, we plot on figure~\ref{fig:lxtrelation} the
$L_X-T$ relation of equation~(\ref{eq:lxt}) together with data taken for
groups \citep[from][]{HP2000} and clusters of galaxies
\citep{Markevitch98,AE1999}. The quality of the fit is obviously
excellent. Note that we adopt for $T_0$ the value computed in
Sec.~\ref{sec:compt0} and used $Q_X = 0.115$. There is thus no free
parameter in the solid curve computation, since $K_0$ has been already fixed.

Note that the hotter groups seem to be overluminous when
compared to the cluster data as well as to the analytic relation. This
is due to the fact that, in the cluster observations, the central
cooling-flows were cut \citep{Markevitch98} or the sample was chosen
explicitly to contain only clusters with known weak cooling-flows
\citep{AE1999}. On the contrary, the groups luminosity in
\citet{HP2000} are not corrected for the cooling-flow (which explains
that the hotter groups, which are expected to have the greatest fraction of their
luminosity coming from their cooling-flow because of their higher central
density, show the greatest departure). Our model does not take into
account a possible cooling-flow component, which explains that it fits
very well the cluster data, while lying near the lower envelope of
the hotter groups. It fits nevertheless very well the lower
temperature groups\footnote{For more discussion on this point, see
\citet{BM2001}, Sec. 2.1.}.  

Both the cluster slope and the group steeper
slope come out naturally of our relation. This agreement is remarkable
and is the first to our knowledge \emph{based on an analytic
discussion to encompass both groups and clusters.} 

This good agreements is a strong sign in favor of the consistency of our
approach. It is thus natural to try to apply
an analogous method to the correlation between the mass of gas and the mean
temperature, as well as to the SZ effect, and compare cluster observations with
our analytic predictions. Unfortunately, gas mass observations are sparse and
heterogenous for groups, while no SZ effect in a group has yet been
observed. We can nevertheless make predictions for the $M_{ICM}-T$ and
$y-T$ relation in this range, which could be validated by undoubtly
forthcoming SZ measurements.   

\section{From the adiabat to the ICM gas mass and gas fraction}
%****************************************************************
\label{sec:fgas}

\subsection{Predicting the $M_{ICM}-T$ relation and comparing to observations}
%-----------------------------------------------------------------------------

Another tight correlation found in reducing ICM data of clusters is the link
between the ICM mass and the mean temperature. For a perfectly self-similar
model, one expects theoretically $M_{ICM} \propto T^{3/2}$. In fact, several
groups have reported recently a steepening of this relation, which could be
due to entropy injection. \citet{VFJ99} have studied the outer regions of a
ROSAT sample of clusters, and found that the above correlation could be
written $M_{ICM} \propto T^{1.71\pm 0.13}$, steeper but nevertheless close to the
self-similar value. On the other hand, \citet{MME99} have fitted double
$\beta$-models to another large sample of clusters and found that $M_{ICM}
\propto T^{1.98 \pm 0.18}$, much steeper than the self-similar model. It is not clear
if this discrepancy (at only $1 \, \sigma$ level) is real and if so, what are
the reasons behind it. 

It is easy to compute this correlation in our framework and
worth comparing our predictions with observations. The ICM mass of gas
 can be written (assuming spherical symmetry)~:  
\be
\label{eq:micm}
M_{ICM} = 4 \, \pi \int^{r_v}_0 r^2 \, \rho_{gas}(r) \, dr,
\en
where $\rho_{gas}(r) = \mu \, m_p \, n(r)$ is the mass density of
plasma. Rescaling with the virial radius and replacing the gas number density
by a combination of temperature and adiabat, one finds~:
\be
\label{eq:micmpred}
M_{ICM} =  M^0_{ICM} \, T^3 \left[1+\frac{T}{T_0}\right]^{-3/2}
\en
with
\be
\label{eq:micmnorm}
M^0_{ICM} = \mu \, m_p \left(\frac{3 \, M_*}{\Delta \, \rho_c^0 \,
T_*^{3/2}}\right) K_0^{-3/2} Q_M,
\en
where we defined naturally~:
\be
\label{eq:qmdef}
Q_M = \int_0^1 \frac{x^2 \, dx}{f^{3/2}}\; .
\en 
The value of this shape factor, which determines the overall normalisation,
is also evaluated in Appendix~\ref{app:shapefactor}. The predicted value
using eq.~(\ref{eq:qmdef}) is $Q_M = 0.1915$, while the best-fit
normalisation is $\sim 40 \, \%$ lower. We argue in
Appendix~\ref{app:shapefactor} that the observed surface 
brightness in most of the  clusters does not reach $r_{500}$ (the virial
radius) and thus needs an risky extrapolation of the data. Moreover, the
observational derivation of $r_{\Delta}$ is also dependent on several
assumptions (in particular on the assumption of a given mass profile) and it
is possible that the gas masses were in fact computed at $\Delta \not=
500$. Indeed, lowering the the upper boundary of the integral in
eq.~(\ref{eq:qmdef}) to 0.8 ({\it i.e.\/,} assuming $r_{500}$ is
underestimated by only $20 \, \%$) gives $Q_M = 0.1136$, which is in perfect
agreement with the data. We use this value in figure~\ref{fig:mgastrelation}
and in the following section.

The slope of the correlation also steepens from very hot clusters ($M_{ICM}
\propto T^{3/2}$, analogous to the self-similar correlation) to groups ($M_{ICM}
\propto T^3$). The temperature range probed by \citet{MME99} is intermediate
between these two behaviours. 

The figure~\ref{fig:mgastrelation} shows the data from \citet{MME99}, as well
as their best-fit linear correlation $M_{ICM} \propto T^{1.98}$ (dot-dashed
line). The solid line is our predicted relation using the computed value $Q_M
= 0.1136$.  The predicted slope matches perfectly the data. The steepening of
the relation compared to the self-similar one is thus, in our particular
model, simply a consequence of the differential shock strength and the
entropy floor. 

%-----------------------------------------------------------
   \begin{figure} 
   \centering 
   \includegraphics[width=9cm]{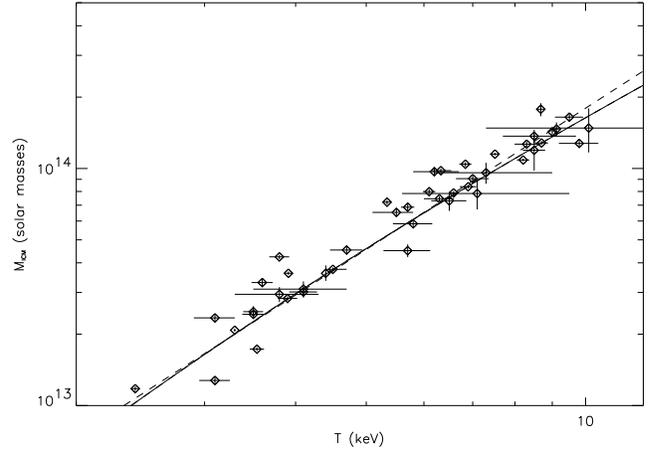}
   \caption{$M_{ICM}-T$ correlation. The data are from \citet{MME99}. The
   dot-dashed line is the best-fit linear correlation $M_{ICM} \propto
   T^{1.98}$ found by these authors, while the solid line is the predicted
   correlation (eq.~\ref{eq:micmpred}) using the computed value $Q_M =
   0.1136$ (see Appendix~\ref{app:shapefactor}). The curvature of our
   prediction is clearly visible when compared to the best-fit power-law.
   \label{fig:mgastrelation}}
   \end{figure}
%-----------------------------------------------------------

\subsection{Comparing the subsequent $f_{gas}-T$ relation with observations}
%--------------------------------------------------------------------------

The gas fraction is very important in that it is supposed to give a direct
lower limit on the universe baryon fraction, if clusters are a fair sample of the
universe \citep{WNEF93}. Such measurements have proven to be tight
constraints to the total mass density of the universe. 

Using eq.~(\ref{eq:micmpred}) and  the observed $M-T$ relation of
eq.~(\ref{eq:mt}), \ie the observed scaling between total mass and
temperature, it is straightforward to compute the gas fraction
variation with temperature~: 
\be
\label{eq:fgast}
f_{gas} = f^0_{gas} \, T^{3/2} \left[1+\frac{T}{T_0}\right]^{-3/2}.
\en
with 
\be
\label{eq:fgasnorm}
f^0_{gas} = \mu \, m_p \left(\frac{3}{\Delta \, \rho_c^0}\right) K_0^{-3/2}
Q_M \: .
\en   
The value of $Q_M$ is the same as in the last section ($Q_M=0.1136$) and no
renormalisation is performed.  

Eq.~(\ref{eq:fgast}) is shown in figure~\ref{fig:fgastrelation},
together with observations. The data points are from
\citet{MME99}. These data were fitted by the authors with a power-law,
namely $f_{gas} = (0.207 \pm 0.011)(T/6\,{\rm keV})^{0.34 \pm
0.22}$. The shaded area, limited in abscissa by the minimum and
maximum temperature of the sample, represents the area covered by similar power-laws
increased and decreased by $1 \, \sigma$ (both in normalisation and in
slope). It is obvious that the data are dispersed, but the trend found by
\citet{MME99}  is very well recovered by eq.~(\ref{eq:fgast}), as well as
the normalisation. 

%-----------------------------------------------------------
   \begin{figure} 
   \centering 
   \includegraphics[width=9cm]{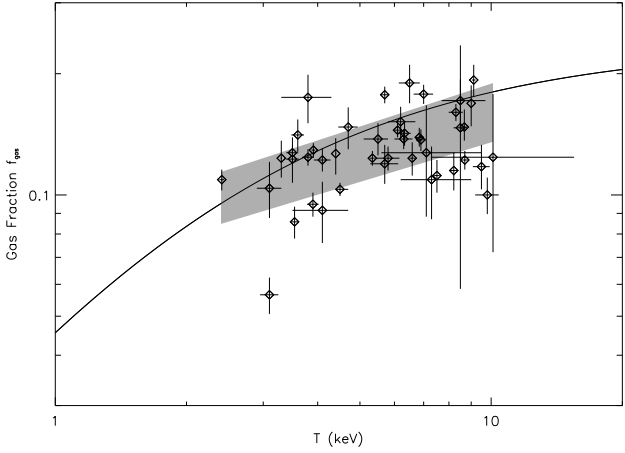}
   \caption{$f_{gas}-T$ relation. Data are taken from \citet{MME99}
   as well as the shaded area which corresponds to the domain of validity of
   their fitted power law, $\pm 1\sigma$. The solid line corresponds
   to our prediction. Both the trend and the normalisation are in good
   agreements. \label{fig:fgastrelation}}
   \end{figure}
%-----------------------------------------------------------

Looking at eq.~(\ref{eq:fgast}), one sees that for $T_0 \ll T$, \ie
for rich clusters, the gas fraction becomes independent of the
temperature. This regime is only reached asymptotically in the figure.
On the other end the slope steepens from clusters to groups, since we have~:
\bea
\label{eq:fgastgp}
f_{gas} & \propto & f_{gas}^0 \, T_0^{3/2} = \quad \mathrm{ constant\quad
for\quad rich\quad clusters,} \nonumber \\
& \propto & M \quad \mathrm{\quad for\quad groups,} 
\ena
where $M$ is the total mass of the group. It is interesting to note that the
fact that the gas fraction is proportional to the total mass in the groups
regime was first claimed by \citet{BBP99}. We thus recover analytically
the adiabatic behaviour modelled by these authors. However, their model
breaks down at high temperature, when the shocks become important, while
ours runs smoothly from an adiabatic infall to a shock-dominated regime.    

\section{From the adiabat to the SZ temperature decrement}
%*********************************************************
\label{sec:yt}

The observations against which our model was validated above were only in the
X-ray wavelengths. We are now turning to the radio and sub-millimeter bandwidth, which
also probes the hot gas in clusters (via the Sunyaev-Zeldovich  effect) and
offers independent observations, with different 
possible systematics and errors. This allows us to independently
validate the model and make some predictions about the group behaviour in
this band.  

Inverse-Compton scattering of incoming CMB photons on ICM thermal
electrons cause a well observed change in the spectral dependence of the
CMB \citep{Re95,Bi99}. The amplitude of the CMB temperature brightness
variation can be written~: 
\be
\label{eq:deltatcmb}
\frac{\Delta T}{T_{CMB}} = h(x) \int \left( \frac{k T}{m_e c^2}\right)
n_e \sigma_T dl \equiv h(x) \, y\: ,
\en
where $h(x)$ is the frequency dependence (with $x = h \nu/k T_{CMB}$), which
simplifies to $h(x) = -2$ at the Rayleigh-Jeans part of the spectrum, 
$y$ being called the comptonization parameter. The integral in the
right-hand side of equation~(\ref{eq:deltatcmb}) can be expressed as a
function of $K$, as for the X-ray luminosity~:  
\bea
\frac{\Delta T}{T_{CMB}} & = & h(x) \, y \\ 
& = & h(x) \frac{k \sigma_T}{m_e c^2} \int n_e T dl \\
& = & h(x) \frac{\sigma_T}{m_e c^2} \int \frac{(k T)^{5/2}}{K^{3/2}} dl \label{eq:yk}
\ena

\subsection{An analytic prediction for the $y-T$ relation}
%---------------------------------------------------------
\label{sec:analyt}

For current SZ experiments resolution (especially interferometric ones), the
beam smearing is still a critical issue as long as we want to determine
a central value. Most authors use X-ray determined temperature as well as the
X-ray surface brightness to correct for the beam and to obtain the 
central value of the SZ effect $y(0)$. This might be a source of
uncertainty but we will nevertheless express this quantity in our physical
framework, and compare it to observations.  

The beam-corrected quantity can be obtained in the following way : going back
to equation~(\ref{eq:yk}) we write (assuming spherical symmetry) 
\bea
 y_0 &  \equiv & y(0) \nonumber \\
& = & \frac{\sigma_T}{m_e c^2} \int \frac{T^{5/2}}{K^{3/2}(T)}
dl \label{eq:yt32} \\
& = & \frac{\sigma_T}{m_e c^2} \frac{T^{5/2}}{K_2^{3/2}(T)}
\int_R^\infty \frac{r dr}{\sqrt{r^2 - R^2} f^{3/2}(r)} \nonumber \\
& = & \frac{\sigma_T}{m_e c^2} \frac{T^{5/2}}{K_2^{3/2}(T)} R_v
\int_0^\infty \frac{dx}{f^{3/2}(x)}\: .
\ena
Using eq.~(\ref{eq:knorm}), (\ref{eq:rvm}) and (\ref{eq:mt}) we have thus~:
\be
\label{eq:y0}
y_0 =  \overline{y}_0 \,\, T^{3} \left[ 1 + \left( \frac{T}{T_0} \right) \right]^{-3/2}
\en
with
\be
\label{eq:ynorm}
\overline{y}_0 = \frac{\sigma_T}{m_e c^2} \left[ \frac{3 M_*}{4 \, \pi \, \Delta \,
\rho_c^0 \, T_*^{3/2}}\right]^{1/3} K_0^{-3/2} Q_{SZ} \: , 
\en
where we have defined~:
\be
Q_{SZ} = \int_0^\infty \frac{dx}{f^{3/2}(x)}\; .
\en
%Note that since in practice the SZ effect is not observed beyond $0.6 \,
%r_v$ and we do not expect much signal outside the virial radius, we limit the
%upper boundary of this integral to 0.6 when evaluating $Q_{SZ}$
%(cf. Appendix~\ref{app:shapefactor}). 
The value of $Q_{SZ}$ is computed in Appendix~\ref{app:shapefactor}.

Finally, let us make a comment about the shape of the $y-T$
relation. For $T \gg T_0$, we get~: 
\be 
\label{eq:ytclus}
y_0 \propto T^{3-3/2} \propto T^{3/2}\: ,
\en
while for $T \ll T_0$ :
\be 
\label{eq:ytgp}
y_0 \propto T^{3}.
\en
Thus, as a matter of fact we expect a steepening of the $y-T$ relation
when going from clusters to groups.

\subsection{Comparing with $y-T$ observations}
%---------------------------------------------

%-----------------------------------------------------------
   \begin{figure} 
   \centering 
   \includegraphics[width=9cm]{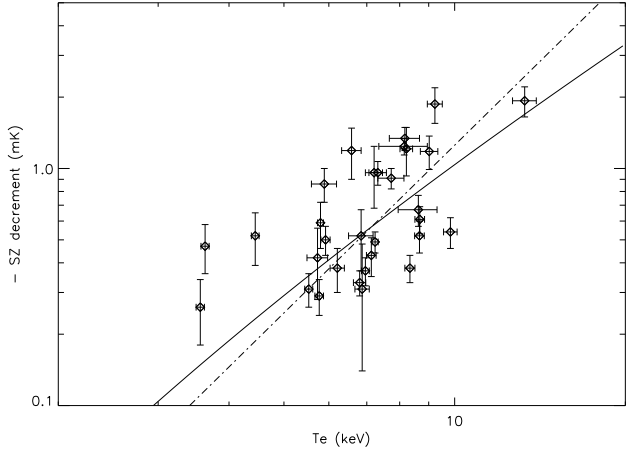}
   \caption{Central SZ decrement - X-ray temperature relation. The data
   points are taken from \citet{Zhang2000}. The solid
   line is the predicted $y_0$ using $Q_{SZ} = 1.835$ . The dot-dashed line
   corresponds to the renormalised correlation fitted by \citet{Cooray1999}.
   The predicted normalisation was computed using  $K_0 = 102 \, \rm keV
   cm^2$ and $T_0 = 2 \, \rm keV$, {\it i.e.\/,} it is \emph{not free} once
   $K_0$ has been fitted to the observations (Sec.~\ref{sec:ktobscomp}).   
   \label{fig:ytrelation}}
   \end{figure}
%-----------------------------------------------------------

Taking observationnal data from 
\citet{Zhang2000} (diamonds), we can compare our predicted relation
of equation~(\ref{eq:y0}) to observed correlations. 

In figure~\ref{fig:ytrelation} we plot 
data and their associated error bars, our prediction for $Q_{SZ} = 1.835$
(solid line, computed in Appendix~\ref{app:shapefactor})  and
the best fit of \citet[][dot-dashed line]{Cooray1999}\footnote{Note that in
\citet{Cooray1999}'s paper, the quoted normalisation of the best-fit $y-T$
relation is wrong by a factor of 10.}. The fact that only rich cluster data
are yet available (which limits the leverage on the slope) and the large
dispersion on the observed y-T correlation translates into a large
uncertainty on the best-fit slope : $2.35 \pm 0.85$.   

Our prediction for the beam-corrected value is steeper than the Cooray's
best-fit relation but still consistent with it at the 1$\,\sigma$
level. It seems to be in better agreement with Zhang's data set,
which extends to slightly higher and lower temperatures than
Cooray's. Note that we will further show that equation~(\ref{eq:y0}) is also
in very good agreement with semi-analytic models for both groups and
clusters. 

The overall agreement with both data sets is less impressive than for the
previously derived $L_X-T$ relation, but is still very reasonnable, given the 
observational dispersion.

Since X-ray and SZ observations are completely independent, the agreement
between our analytic model with the latter is another independent
confirmation of its validity. A robust confirmation would
come from SZ observations of groups of galaxies (unfortunately not yet
feasible) and from the change in the correlation slopes. Nevertheless, we
can compare our simple model to more elaborate semi-analytic models and their
predictions. 

Menci \& Cavaliere (2001) have presented semi-analytic predictions for the
observed SZ effect in groups and clusters. Their semi-analytic model takes
into account the preheating due to star formation, in order to bend the
$L_X-T$ relation in the groups mass range. Our predictions for the $y(0)-T$
relation (namely equation~(\ref{eq:ytclus}) for clusters and (\ref{eq:ytgp})
for groups) are in complete accord with their preheated models. They found a
self-similar relation for high mass systems ($y \propto T^{3/2}$) which is
exactly our prediction. The slope in the groups mass range depends on the
feedback model they chose, but our prediction ($y(0)\propto T^3$) is
also similar to their computed moderate feedback model (which looks to
be a better choice, since their strong feedback model is likely to give only an upper
limit on the feedback effect). We are thus able to reproduce their figures 3
and 4 with our simple analytic model. Even if this is not a confirmation of
the model (since we use an analytic version of their model of shocks and our
hypothesis on the preheating are very similar), we find it extremely
satisfying to reproduce {\it analytically}, and in a physically
straightforward and motivated way, 
the results of a more complicated and non-analytic model. This gives us more
confidence in the fact, that we have captured in this simple physical scheme
the essential ingredients of the clusters and groups formation.

\section{Linking SZ and X-ray measurements}
%=========================================
\label{sec:ylx}
%-----------------------------------------------------------
   \begin{figure} 
   \centering 
   \includegraphics[width=9cm]{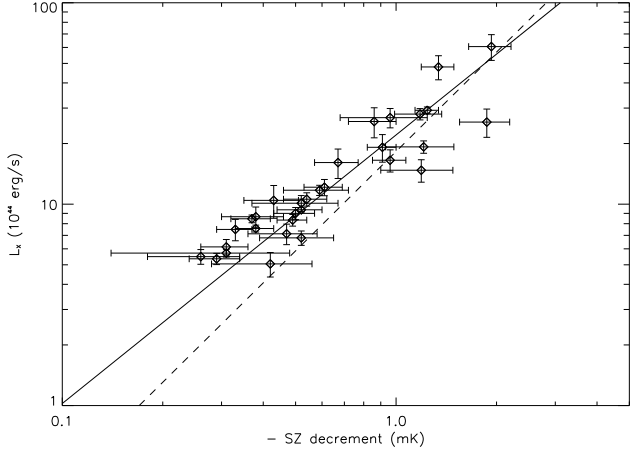}
   \caption{Central SZ decrement - X-ray luminosity relation. The solid line
   represents the analytic predicted correlation, using the predicted values
   of $Q_{SZ}$ and $Q_X$ found in Appendix~\ref{app:shapefactor}. The dashed
   line corresponds to \citet{Cooray1999}'s best-fit correlation. The data
   are taken from \citet{Zhang2000}.}   
   \label{fig:ylxrelation}
   \end{figure}
%-----------------------------------------------------------

To link SZ and X-ray measurements, it is natural to determine a
correlation between $y$ and $L_X$. It is expected to be much tighter than the $y$-$T$
relation \citep[see][]{Cooray1999}, but is more intricate to obtain
analytically. In this section, we will give two analytic expressions for
this correlation in the rich clusters's and in the groups' range mass respectively.

% Using equations~(\ref{eq:ybeam}) and
% (\ref{eq:lxt}), we can write~:
% \be
% \label{eq:lxyt}
% \frac{L_X}{\overline{y}_{\theta_b}^2} \propto T^{\alpha - \xi - 2}.
% \en
% We need now to invert equation~(\ref{eq:ybeam}) but this can not be
% easily done without new hypotesis. Since the SZ observations span only the
% range of clusters yet, let's assume that $T\ggT_0$. We can then write~:
% \be
% T \propto \overline{y}_{\theta_b}^{2/5},
% \en
% and inserting this in equation~(\ref{eq:lxyt}) gives~:
% \bea
% L_X & \propto & \overline{y}_{\theta_b}^{(6+2(\alpha-\xi))/5} \\
%  & \propto & \overline{y}_{\theta_b}^{4/5} \, \, \mathrm{with} \,\,  \xi=3/2 \,\,
% \mathrm{and} \,\, \alpha = 1/2
% \ena
Indeed using equations~(\ref{eq:y0}) and (\ref{eq:lxt}), we can write~:
% SERGIO 
%\be
%\label{eq:lxyt}
%\frac{L_X}{y(0)} \propto T^{\alpha + \xi/3 - 2}.
%\en
\be
\label{eq:lxyt}
\frac{L_X}{y_0^2}  = \frac{L_X^0}{\overline{y}_0} \,  T^{\alpha-3/2}.
\en
To proceed further we need now to ``invert'' equation~(\ref{eq:y0}) but this can not be
easily done without a new hypotesis. Since the SZ observations span only the
range of clusters yet, as long as we want to compare to observations
it is satisfying to assume that $T \gg T_0$. We can then write~:
\be
T  =  T_0^{-3/2} \left(\frac{y_0}{\overline{y}_0}\right)^{2/3},
\en
and inserting this in equation~(\ref{eq:lxyt}) gives~:
\bea
L_X & =  & L_X^0 \, T_0^{3/2} \left(\frac{y_0}{\overline{y}_0}\right)^{1+\frac{2}{3}\alpha} \\
& = & L_X^0 \, T_0^{3/2} \left(\frac{y_0}{\overline{y}_0}\right)^{4/3} \,\, \mathrm{with} \,\, \alpha = 1/2. \label{eq:ylx}
\ena
The observed correlation between $y$ and $L_X$, together with the analytic
prediction (equation~(\ref{eq:ylx}), solid line) is shown in
figure~\ref{fig:ylxrelation}. The data are taken 
from \citet{Zhang2000} and the dashed line is the best-fit relation found by
\citet{Cooray1999}. The normalisation has been computed using $K_0 = 102 \,
\rm keV cm^2$, $T_0 = 2 \, \rm keV$, $Q_X = 0.115$ and $Q_{SZ} = 1.835$.  The
agreement is very good, even if it must be recalled that 
the shown correlation has been computed for the case $T \gg T_0$. Thus the real
correlation will be slightly steeper (thus more in agreement with Cooray's
correlation), but by a very small amount, well within the observational
dispersion.  

Finally, one can compute this correlation in the groups mass range ($T \ll
T_0$), yielding a steeper correlation~:
\be
\label{eq:ylxgp}
L_X = L_X^0 \left(\frac{y_0}{\overline{y}_0}\right)^{3/2} \,\, \mathrm{with} \,\, \alpha = 0. 
\en

%*********************************************************
\section{Discussion}
%*********************************************************
\label{sec:disc}

 We have shown that a simple modeling of the main process in charge of the
 formation of clusters (namely the shocks which thermalise the gas inside the
 virial radius) allows to derive a general $K-T$ relation that reproduces
 very well the observed one. Moreover, we were able to deduce most of the
 observed scaling laws with the adequate normalisation, encompassing both
 groups and clusters. Thus, the physical meaning of \eg the $L_X-T$ and $y-T$
 relations, and in particular of the break in their self-similarity when
 going from clusters to groups (only evidenced in X-rays for the moment) can
 be understood as a relic of the formation process of these systems~: a
 competition between an entropy floor prior to the shock and the shock
 process itself. This suggests naturally that entropy constitutes the best
 ``observable'' in clusters, as already emphasized by the modified spherical
 model of \citet{TN2001}. What we add here is that we show that entropy (in
 fact, the adiabat) is also suitable to seek analytic expressions for the
 different correlations found observationally in groups and
 clusters. Reversing this argument, these scaling laws
 should be an appropriate probe of the entropy content of clusters.

 \subsection{The $L_X-T$ relation as a probe of entropy content}
%--------------------------------------------------------------
 \label{sec:entropcont}

 While the entropy spatial variations of the baryonic gas in clusters is a
 valuable piece of information on their present thermodynamic state \citep[in
 particular for merging clusters, see {\it e.g.\/,}][]{MSV99}, the total
 entropy content $\overline{K}$ (hereafter TEC), defined as
 \be
 \overline{K} = \int_V K({\mathbf r})\, dV
 \en
 is related to the integrated thermodynamic history of the formation of these
 systems. Therefore, any non-gravitational entropy injection at some point in
 this history will enhance the entropy content over the shock-generated one,
 while the cooling should decrease it in the central parts.

 The analytic model we have presented in the last sections allows us to shed
 some light in the TEC and its link to the $L_X-T$ relation. On the one hand,
 given our assumption of isothermality, eq.~(\ref{eq:intk3}) indicates that
 $L_X/T^{3+\alpha}$ is related to the TEC in a given system. On the other
 hand, the self-similar models of cluster formation \citep{Kaiser86} predict
 that $L_X\propto T^2$, which gives directly (using $\alpha = 1/2$)~:
 \be
 \label{eq:tecss}
 \frac{L_X}{T^{7/2}} \propto T^{-3/2}.   
 \en
  Any departure from this last relation witnesses entropy injection or
 loss. Consequently, at a given temperature, the gap between the curve
 representing eq.~(\ref{eq:tecss}) and the locus of observed groups and
 clusters will give a indirect measure of entropy gain at a given mass (if
 the central cooling parts are adequately excised). Unlike the observations
 of the ``entropy floor'', which were obtained at a single scaled radius near
 the center \citep{PCN99,LPC2000} and so are representative of a particular
 point in the entropic history ({\it i.e.\/,} at a time near the last major
 merger) , \emph{this measure includes most of the volume} of the systems and
 thus most of its past thermodynamic history. However, this does not directly
 measure $\overline{K}$, but the integral of a power-law of the TEC. We will
 first show that in a semi-analytic scheme (with less assumptions than the
 present analytic model as well as with a realistic entropy evolution of the
 intergalactic medium) the interpretation of the $L_X-T$ relation is still
 valid and then discuss how X-ray and SZ observations can constraint the TEC.

 \subsection{Relaxing some hypotheses}
 %--------------------------------------------
 \label{sec:semianal}
 
 Assuming local thermodynamic equilibrium, isothermality, and the
 self-similarity of the adiabat profile, we illustrated how the observed
 $L_X-T $ relation is a probe of the TEC. Whereas the LTE is quite robust
 (and actually very difficult in practice to alleviate), neither the
 isothermality nor the self-similarity are fully exact from both an
 observational and a theoretical point of view. In particular, this last
 assumption is a very strong one and will be discussed in depth in
 section~\ref{sec:ssassum}. It is thus worth relaxing these hypotheses and
 testing the validity of our work by considering how well the quantity
 $\int_V\ K^{-3} dV \displaystyle$ probes $L_X/T^{7/2} \displaystyle$, \ie
 testing eq.~(\ref{eq:intk3}) when we consider a full non-isothermal and
 non-self-similar profile.

 To this purpose, we will use a semi-analytic model (hereafter SAM)
 developped independently by one of us (Dos Santos 2001, in
 preparation). This particular model uses the conduction-structured
 temperature profile (which was shown by \citet{SSCD2001} to describe very
 well the temperature and surface brightness profile of clusters of
 galaxies), together with an NFW profile for the dark matter density
 profile. A shock model at the virial radius, together with entropic
 constraints at the center allow to predict the temperature profile without
 the hypothesis of isothermality or the unphysical polytropic link between
 temperature and density. The evolution of the central entropy is governed by
 the entropy evolution of the IGM in the universe obtained from the model of
 \citet{VS99} in two different cases~: in the first one, the reheating is
 provided by SN explosions only, while in the second one, AGN and quasar
 activity provide the entropy injection. Both cases were validated against a
 number of observations, including the $L_X-T$ relation, the change in
 surface brightness profiles from clusters to groups, the baryon fraction in
 these systems and the entropy floor.

 Figure~\ref{k3dvtrelation} shows the comparison between observations and the
 SAM. The solid (SN case) and dashed lines (QSO case) show the quantity
 $\int_V\ K^{-3} dV \displaystyle$ computed directly using the model. Since
 the specific entropy is known up to an additive constant, both lines were
 normalised to match the high temperature clusters. The dot-dashed line shows the
 self-similar prediction for $L_X/T^{7/2} \displaystyle$
 (eq.~\ref{eq:tecss}), also renormalised to match rich clusters. The
 preheated models high temperature slopes match naturally the self-similar
 prediction. They also match very well the cluster data and their trend in
 temperature. It is in particular remarkable that the three clusters with the
 smallest temperatures ($T \sim 2 \, \rm keV$) depart notably from the
 self-similar prediction, and lie exactly on top of the semi-analytic
 prediction. The lower envelope of groups is well followed by the SAM, while
 some points are over this prediction. We think that, as in
 Sec.~\ref{sec:lxtobs}, this is due partly to the fact that the central
 cooling regions of groups were not removed, unlike the clusters, and partly
 to an intrinsic scatter.

 %-----------------------------------------------------------
    \begin{figure} 
    \centering 
    \includegraphics[width=9cm]{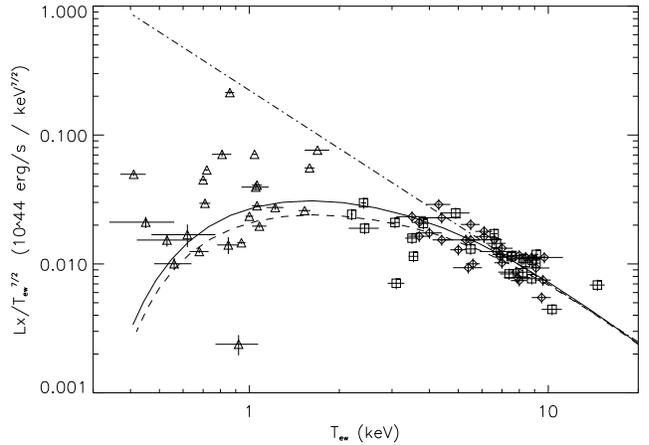}
    \caption{$L_x/T^{7/2}$ (observational points) compared to the direct
    computation of $\int_V K^{-3}({\mathbf r}) \, dV \displaystyle$ in the
    SAM. The data are the same as in the figure~\ref{fig:lxtrelation}. The
    solid (respectively. dashed) line is the result of the SAM in the case of
    SN (respectively. QSO) reheating. The dot-dashed line is the self-similar
    prediction ($L_X/T^{7/2} \propto T^{-3/2} \displaystyle$). Thus, relaxing
    the assumptions of isothermality and self-similarity of our analytic model
    does not change the interpretation of $L_X/T^{7/2} \displaystyle$ as a
    probe of $\int_V K^{-3}({\mathbf r}) \, dV \displaystyle$. This explains
    that our simple model is able to recover the slope and the normalisation
    of the $L_X-T \displaystyle$ relation. 
    \label{k3dvtrelation}}
    \end{figure}
 %-----------------------------------------------------------

 The figure shows why our simple model works well in recovering slope and
 normalisation of the $L_X-T$ relation (and of the other relations we
 studied) : the equality~(\ref{eq:intk3}) is also verified in a more general
 semi-analytic model where the adiabat profile is fully known and the ICM is
 not isothermal. At a given temperature, the difference in location between
 observations and the self-similar curve will give a measure of the
 non-gravitational entropy injection in systems with this mean
 temperature. \emph{This measure does not depend at all on hydrostatic
 equilibrium or spherical symmetry. Nor does it require either that the
 preheating was external.} In fact, eq.~(\ref{eq:lxt3}) is completely
 independent of any modeling of the entropy injection. It only requires
 isothermality and local thermodynamic equilibrium on the scales probed by
 the observations. It thus can be used as a powerful, model-independent,
 constraint on all the internal or external preheating models, as well as on
 differential galaxy formation efficiency between groups and clusters
 \citep{Bryan2000}. Unfortunately, the TEC is not measured directly, but the
 integral of the right-hand side of eq.~(\ref{eq:intk3}) will obviously
 decrease as the entropy injection amount is increased. Thus, if the adiabat
 profile is known in a given model, it will be easy to relate this integral
 to the TEC, even if a certain amount of degeneracy will obviously be present
 (this measure being integrated in space and time, different entropy
 injection histories can lead to the same final amount of entropy at $z=0$).

 In fact, this constraint will be more or less the same as the constraint
 given by the $L_X-T$ relation (already used to constrain the models), even
 if the interpretation in terms of entropy is physically more satisfying.
 However, the same remarks can be made with the surface integrated SZ
 decrement. Indeed, the SZ y-Compton parameter integrated over the whole
 surface of a cluster can be written~:
\be 
 \label{eq:yint}
 \frac{y}{T^{5/2}} = \frac{\sigma_T}{m_e \, c^2} \int_V
 \frac{dV}{K^{3/2}({\mathbf r})}, 
 \en
 which shows that the quantity $y/T^{5/2} \displaystyle$ is also a probe of
 the entropy content, but with a different power of the adiabat profile
 inside the integral. Thus, once SZ observations for groups are available,
 even non-resolved, the combination of X-rays and SZ at a given temperature
 will be a powerful constraint on cluster formation models. Note that the
 relation $M_{ICM}-T$ gives an analogous relation as eq.~(\ref{eq:yint}), and
 can thus be used now as an entropy probe.

\subsection{The adiabat profiles and the self-similarity}
\label{sec:ssassum}

\begin{figure}
\label{fig:scaledkprof}
\includegraphics[width=9cm]{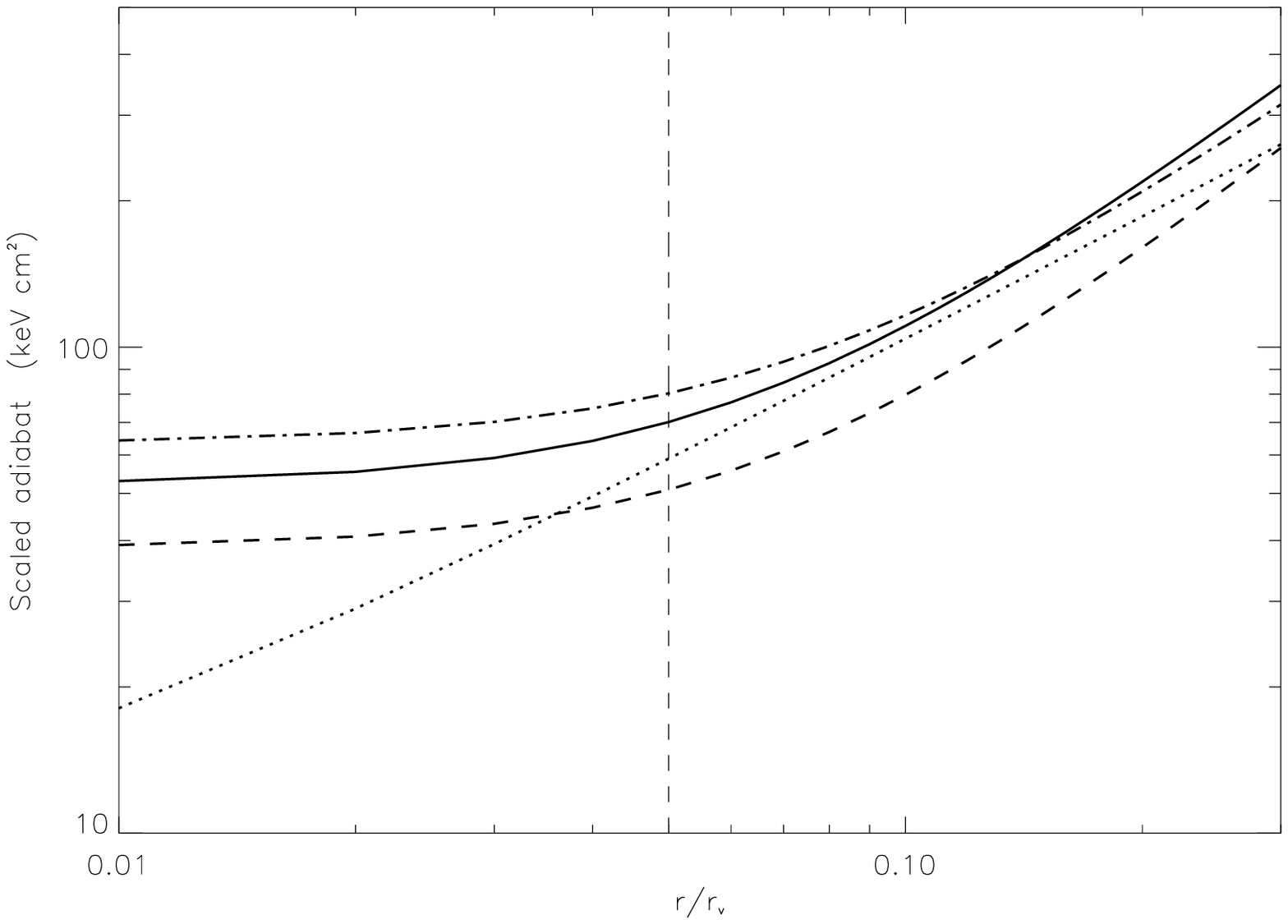}\hfill
\includegraphics[width=9cm]{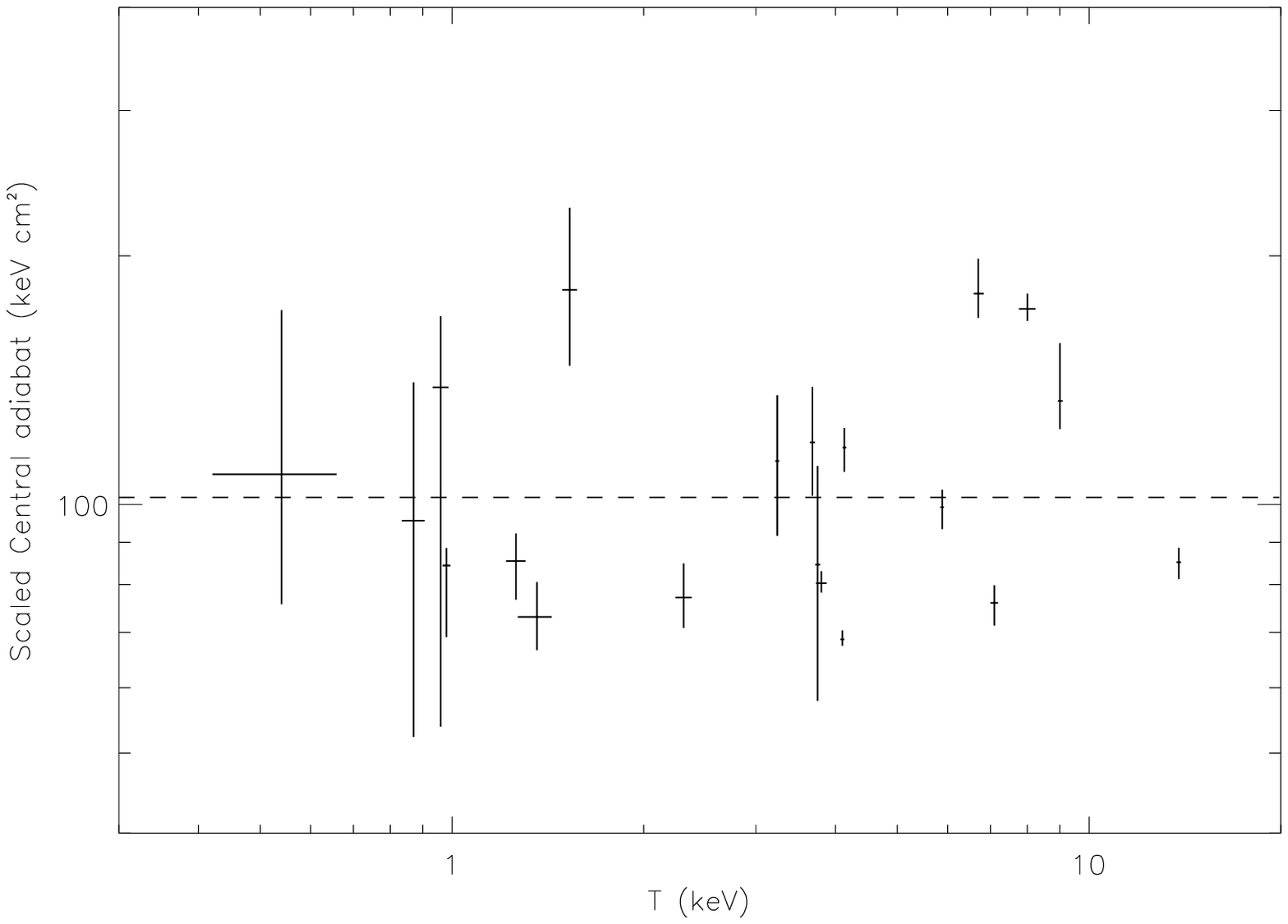}
\caption{{\it Upper panel~:} Average scaled adiabat profiles for systems
grouped by mean temperature (solid~: 6-14 keV; dashed~: 3.7-6 keV;
dot-dashed~:  1.3-3.7 keV; dotted~: 0.5-1.3 keV). Each individual profile was
obtained from the best-fit gas and density profiles given by \citet{LPC2000}
and scaled by $\left[1+T/T_0\right]/(1+z) \displaystyle$. No trend is found
with the temperature  \citep[unlike with the scaling used by][]{LPC2000} and
all the profiles are approximately self-similar. The vertical dashed line
shows the mean 
cooling-flow radius. {\it Bottom panel :} Scaled central entropy value
as a  function of the mean temperature of each system. Again, no excess in
groups is found, which validates our assumption of self-similar adiabat
profiles (at least in the core) and our approach. The dashed line shows the
entropy floor value infered from the data ($K_0 = 102 \, \rm keV \,
cm^2$, corresponding to $K_1 = 120 \, \rm keV \, cm^2$).}   
\end{figure}
 To derive the analytic scaling relations presented in this paper, we have
 used a strong assumption : the self-similarity of rescaled adiabat profiles
 (see eq.~\ref{eq:kprof}) {\it i.e.\/,} the fact that the temperature enters
 only the normalisation of the adiabat profile. The validity of this
 assumption is questionable, both on observational and theoretical grounds.

 \citet{PCN99} show that density profiles are shallower in groups than in
 clusters \citep[even if this result is still uncertain, see][for an
 alternative view]{RSB2000}, while preheating models predict naturally that
 entropy gradients in lower temperature systems are smaller than in clusters
 \citep[due to adiabatic infall during their lifetime,
 see][]{BBP99,TN2001}. However, these models predict also large temperature
 gradients in groups which are not observed \citep{TN2001}. Moreover, using
 1D hydrodynamic models with internal or external preheating, \citet{BM2001}
 obtain linearly rising entropy profiles in groups (outside an isentropic
 core for external preheating models) as well as in clusters. They thus
 produce naturally self-similar adiabat profiles in groups and clusters. The
 reasons of such discrepancies are unclear, but are certainly linked with
 widely differing simulation methods. Note however that \citet{KP97}, using
 similar 1D hydrodynamic simulations, have found shallower slopes in groups
 compared to clusters, \emph{without taking into account any preheating}
 (whose effect is expected to widen the slope difference).

 Spatially-resolved spectroscopic observations of groups and clusters provide
 directly adiabat profiles in groups and clusters and can then help to settle
 this debate. Unfortunately, the low surface-brightness of groups allows this
 type of study only to small radii\footnote{Note that this situation will be
 much improved with groups observations with XMM-Newton, due to its enhanced
 sensitivity.}. Indeed, \citet{LPC2000} obtained the adiabat profile for 12
 systems with $T < 4 \, \rm keV$ up to $r = 0.25 \times r_v$. Even if this
 maximum extent radius is small, the study of the adiabat profiles behaviour
 with temperature inside it is still worth, since the X-ray luminosity comes
 mainly from this central part of the cluster (due to the $n^2$ scaling of
 $L_X$). These authors have rescaled their observed entropy profiles by a
 factor $T^{-1}(1+z)$, removing the effect of system mass and of the
 evolution of the mean density of the universe. They found that the scaled
 adiabat profiles do not coincide, the less massive systems having higher
 scaled entropy profiles (see their figure 4). However, this particular
 scaling does not take into account the physical processes that change the
 adiabat normalisation with mass. What we have shown in this paper is that
 the adiabat profiles of groups and clusters should be rescaled ({\it
 i.e.\/,} divided) by the quantity~:
  \be
\label{eq:scalefactor}
\left[1+\frac{T}{T_0}\right] \, (1+z)^{-1},
\en 
where the $(1+z)^{-1}$ factor
comes from the redshift dependence of the temperature ($T\propto (1+z)$) and
the redshift dependence of the central density (proportional to the mean
density of the universe {\it i.e.\/,}
$\propto(1+z)^3$)\footnote{\citet{LPC2000} do not take into account the
temperature redshift dependence since they divide by $T$.}. 

Using the best-fit $\beta-$model for the gas density and the best-fit linear
ramp for the temperature profile, we reproduced their adiabat profiles, and
rescaled them individually by the quantity in
eq.~(\ref{eq:scalefactor}). Then we grouped the systems by temperature and
computed the mean scaled profile, as in \citet{LPC2000}. The scaled profiles
are displayed in the upper panel of figure~\ref{fig:scaledkprof} \citep[we
used the same line styles as][]{LPC2000}. No trend with temperature is
visible now and the slight dispersion can be attributed to the assumption
that the observation time equals the formation time of a system. Moreover,
the four profiles are very similar between $r=0.05\times r_v$ (the mean
cooling-flow radius, inside which entropy loss is achieved by cooling) and
the outer radius. This means that our assumption of scaled self-similar
adiabat profiles is indeed right, at least in the core of the systems. To
confirm this, one can do the same scaling with the central adiabat values (at
$r = 0.1\times r_v$), which is shown in the bottom panel of
figure~\ref{fig:scaledkprof}. Instead of finding an excess for $T < 4 \, \rm
keV$ as in the figure 5 of \citet{LPC2000}, we find now that all the systems
lie on the same central adiabat symbolised by the dashed line and giving
directly the value of $K_0$ in our model ({\it i.e.\/,} $K_0 = 102 \, \rm keV
\, cm^2$). 

This shows that, \emph{when rescaled conveniently, the
central adiabat profiles of groups and clusters are indeed self-similar} and
thus that our assumption is valid, as well as our whole approach of taking
into account not only hydrodynamic shocks but also an entropy floor
established before them. This explains why our slope and normalisation work
so well for the $L_X-T$ and $y_0-T$ relations, which put more weight on the
core of the systems. It also enlights the fact that our shape factor for the
gas mass is less accurate, since most of the mass lies at large radius where
the self-similarity assumption breaks down (see
Appendix~\ref{app:shapefactor}).

\subsection{Comparison with other results}
\label{sec:compmodels}

We have compared most of our analytic results with predictions from
semi-analytic models and found good agreement (in particular for the trends
of the gas fraction with total mass and the SZ effect with temperature in low
temperature systems). Let us now compare the value of the entropy floor we
need to reproduce the curvature of the $L_X-T$ relation with values assumed
in other theoretical models of energy injection. \citet{BBP99} assume a
constant entropy injection value of $\sim 350 \, \, \rm keV \, cm^2$ and
\citet{TN2001}'s model needs a value in the range $190-960 \, \, \rm keV \,
cm^2$ in order to steepen $L_X-T$. These values, as noted by \citet{LPC2000},
are higher than the observed value, which are likely to be upper limits. On
the other hand, our model needs a value of $K_0 = 102 \, \, \rm keV \, cm^2$,
well within the observational range of $70-140 \, h_{50}^{-1} \, \, \rm keV
\, cm^2$. 

As we have indeed fitted this value to the observations, it may be
asked if we do not force in fact this agreement. The answer is clearly no,
as, first, the agreement of the $K-T$ relation with the observations does not
garantee that the same value of $K_0$  will provide the a good description of
the other relations (both their shape, via the temperature at which the
self-similarity is broken, $T_0$, \emph{and} their normalisation) which range
from radio to X-ray data. Second, the other models reproduce as well the
$K-T$ relation, together with the $L_X-T$ one \citep[at least]['s
model is compared to these observations]{TN2001} \emph{with the same high
value of the entropy floor}. A lower value compatible with observations would
not fit this relation. 

The main difference between our present work and
these models is the fact that their low temperature systems are entirely
isentropic {\it i.e.\/,} no shocks occur at all in groups. This is the
claimed approximation made by \citet{BBP99}, who want to investigate a
limiting model (which naturally fails also in the clusters' mass
range). However, \citet{TN2001} modelise shocks and adiabatic infall, and it
is not clear why shocks do not raise the adiabat profiles of their
groups. 

This clearly shows that \emph{shocks can not be completely
supressed in groups} unless an unacceptably high entropy floor is needed in
order to break the self-similarity in the $L_X-T$ relation. Our model takes
shocks and the entropy floor into account by construction both in clusters
and groups and is thus able to reproduce nicely all the X-ray correlations
with the observed value of $K_0$.

%=========================================================================
\section{Conclusion}
%=========================================================================
\label{sec:conc}

Throughout this paper, we have shown that the adiabat constituted a ``key
observable'' in the ICM, not only because it is a record of the thermodynamic
past history of diffuse baryons in groups and clusters, but also because it
allowed us to derive for the first time \emph{unifying analytic expression
for the slope and normalisation of observed correlations of groups and
clusters} in X-rays and SZ. We have thus derived analytic expressions and
their normalisation for the $L_X-T$, $M_{ICM}-T$, $f_{gas}-T$, $y_0-T$ and
$y_0-L_X$ correlations (for the latter, two limiting expressions have been
provided in the case of groups and of clusters respectively). The
Appendix~\ref{app:a} summarizes these correlations and gives the
corresponding numerical values of their normalisations. We will conclude by
discussing some applications of the model.

Clusters of galaxies, and in particular their baryonic content observed
through its X-ray emission, have often been used to provide cosmological
constraints \citep[][and references
therein]{Pe80,Kaiser86,OBB97,BSBLD2000}. These studies have shown that the
cluster population can provide tight constraints, but have also revealed two
problems (apart from purely statistical problems due to large error bars of
{\it e.g.\/,} the temperature in high-redshift clusters) : first, the
difficulty to model the luminosity in a cluster, mostly due to the fact that
it depends on the core properties, whose formation is still
uncertain. Second, the small available number of rich clusters, in particular
at high redshift, which precludes yet a reliable use of these
tests\footnote{It is paradoxical but amusing to note that this same small
number of high redshift clusters becomes a powerful test of the total density
of the universe \citep{DVGLHS98}.}.

We think that the present work, together with the new generation of X-ray
observatories, can alleviate both of these problems. First, our modelisation
of the cluster luminosity does rely on the physical processes being at work
during the formation of a cluster, and not on an hypothetical density
profile. Therefore, the luminosity does not depend on an arbitrary core
radius and is thus more reliable than previous attempts \citep[in particular,
we don't have to assume anything on the core radius evolution like {\it
e.g.\/,}][]{RVLM2001}.  Second, we extend the analytic relations to the
groups mass scale, taking into account the competition between an external
entropy floor and hydrodynamic shocks (thus recovering the curvature of the
$L_X-T$ relation) and make predictions for other properties of groups. This
is useful since the group population is thought, in a hierarchical structure
formation cosmology, to be much larger than rich clusters. However, its
redshift evolution has never been used to constrain cosmology, partly because
of the lack of reliable data and partly because no analytic modeling of the
groups baryonic properties was available (extrapolating the cluster baryonic
properties was obviously wrong, as the recent observations show). The first
problem will be solved, at least locally, by Chandra and especially
XMM-Newton. Our model provides an attempt to solve the second problem. Of
course, groups will certainly be much more dispersed than clusters (because
the entropy injection will have more effect on them, and any spatial
variation of this injection level will affect the properties of a given
group), but the number of groups detected will be large~: \citet{RVLM2001}
have estimated that more than 100 clusters with $T > 2 \, \rm keV$ will be
detected in a serendipitous survey for $z<0.2$ (irrespective of the
cosmology), a number 10 times highers than for cluster with $T > 4 \, \rm
keV$ and 50 times highers than for $T > 6 \, \rm keV$ for a survey surface of
$\sim 800 \deg^2$. The number of groups with $T < 2 \, \rm keV$ will
obviously be superior, certainly not by a factor 10 because of the reduced
size and steepening of the $L_X-T$ relation, but at least by a factor of a
few. This number of well observed groups (together with pointed observations)
will allow an unprecedent local calibration of the group ensemble
properties. Using our model, it will be possible to use groups and clusters
to constrain cosmology and the cluster LF, TF and $L_X-T$ evolution to add
further constraints. Moreover, groups will provide useful constraints on the
total amount of reheating the present universe underwent and its redshift
evolution if high redshift groups are available (see Sec.~\ref{sec:disc}).

\begin{acknowledgements}
We are deeply grateful to Romain Teyssier whose critical reading
improved significantly the quality of an earlier draft of this paper. S. Dos
Santos aknowledges also very enlightening discussions with
G. C. Stewart. Useful comments from F. R. Bouchet and Y. Mellier are also
gratefully aknowleged.  
\end{acknowledgements}
%************************************************************************

\bibliographystyle{apj}
%\bibliography{masterszx}
\bibliography{papersub}

\appendix

%=========================================================================
\section{Numerical values for the correlations}
%=========================================================================
\label{app:a}

In the following appendix, we provide numerical expressions for the
normalisations of the different correlations the present model predicts.
These will thus be straightforward to use in another context.  The
temperature $T$ is in keV, and the normalisations were computed using a local
Hubble constant of $H_0 = 100 \, h_{2/3} \, \rm km \, s^{-1} \, Mpc^{-1}$
with $h_{2/3} = 2/3$ . We included the dependence on $h_{2/3}$ in each of the
following expressions\footnote{We thus have $h_{50} = 4/3$ and $h_{100} =
2/3$ where $h_{50}~=~(H_0/50)$ and $h_{100}~=~(H_0/100)$.}.

We first describe one relation (external to the model and independent from
it), the $M-T$ relation, that we used to relate the virial radius to
the temperature. Then, we write the different
correlations presented in this work, namely $L_X-T$, $M_{ICM}-T$,
$f_{gas}-T$, $y_0-T$ and $y_0-L_X$.

\subsection{The Mass-Temperature relation}

This relation is crucial, in that it allows us to relate the virial radius
(where the shock is assumed to take place) with the temperature of the
system, via the definition of $M$ in terms of $r_v$. The simulation results
of \citep{EMN96} are frequently used for mass-temperature scaling, but they
seem to provide systematically a higher normalisation than the observed one,
whatever the method used to measure $M$ \citep[][and references
therein]{NMF2000}. We used the observational results of \citet{NMF2000},
derived from observed density and temperature profile, because their sample,
although small, goes all the way from groups to clusters. These authors give
the $M-T$ relation at different scaled radii, from $r_{2000}$ to $r_{500}$
(where $r_\Delta$ is the radius whose mean interior density is $\Delta$ times
the critical density). Since \citet{EMN96} have shown that, inside a radius
$r_{500}$, the baryonic gas is in hydrostatic equilibrium to a very good
approximation and that hydrostatic masses measured within this radius should
be reliable, we choose this radius to normalise the $M-T$ relation. Moreover,
while at smaller radius \citet{NMF2000} find that the $M-T$ relation is
significantly steeper ($99.99 \, \%$ confidence at $r_{1000}$) than the
self-similar prediction ($M \propto T^{3/2}$), the slope they extrapolate to
$r_{500}$ is consistent with 3/2 ($\chi^2/\nu = 5.1/5$, but note that the best
fit has a slope of 1.84). We thus used throughout the paper the following
relation~:
\bea
M_{500} & = & M_* \left( \frac{T}{T_*}\right)^{3/2} \nonumber \\
        &=  & 9. 6  \times 10^{14} \, M_\odot \, h^{-1}_{2/3} \left(\frac{T}{10 \,
\rm keV}\right)^{3/2} \, , \label{eqapp:mt}
\ena
which defines the values $M_*$ and $T_*$ used in the text.
Each time we speak about the virial radius in this paper, we refer to
the radius $r_{500}$ related to $M_{500}$ defined above, and we will thus
take $\Delta = 500$ throughout. 

Finally, let us state that introducing a steeper slope for the $M-T$ relation is
straightforward within the physical scheme we used in this work, but does not
change much the predicted relations gathered in this appendix and their
agreement with observations. Since we wanted to discuss the generic effect of
shocks and preheating, the same effects which are thought to steepen the $M-T$
relation, we found natural to keep a self-similar scaling.

\subsection{The predicted X-SZ-T correlations}

%DEFINITION DE T0

We now compute the different normalisations given in
eq.~(\ref{eq:lxnorm}), (\ref{eq:micmnorm}), (\ref{eq:fgasnorm}),
(\ref{eq:ynorm}) and (\ref{eq:ylx}). For this purpose, we will use the values
of $M_*$, $T_*$ and $\Delta$ from the
last section. The critical density is defined as usual as~:
\be
\label{eqapp:critdens}
\rho_c^0 = \frac{3\,H_0^2}{8\, \pi \, G} \:,
\en
and, for the luminosity, we will need the normalisation of the cooling
function, taken from \citet{ENF98}~: 
\bea
\label{eqapp:coolfunc}
\Lambda_e(T) & = & \Lambda_0 \, T^{1/2} \nonumber \\
 & = & 1.2 \times 10^{-24} \, T^{1/2} \, \, \rm erg \, \, cm^3
 \, \, s^{-1} \:,
\ena
where $T$ is in keV.

Finally, we take for $T_0$ the value computed from $K_0$  in the
Sec.~\ref{sec:compt0} {\it i.e.\/,} $T_0 = 2 \, \rm keV$ and the values of
the shape factors $Q_X$, $Q_{SZ}$ and $Q_M$ computed in Appendix
\ref{app:shapefactor}. \emph{Therefore, the only adjusted quantity used to produce
these relations is $K_0$}, which was fitted to the data of \citet{HP2000} in
Sec.~\ref{sec:ktobscomp} to find $K_0 = 102 \, \, \rm keV \, cm^2$. Once this
quantity was fixed by comparison with the central entropy observations,
\emph{no renormalisation of the following relations is allowed}.  
 
All the numbers are computed here at $z=0$, but it is easy, within a given
cosmology, to extend the formulae in the text to higher redshift (note that
the value of $K_0(z)$ must be assumed as well). The analytic correlations
found are then  

\begin{description}

   \item[$\bullet$] {\bf $\mathbf{L_X-T}$ relation :}

\be
L_X  =  5.643 \times 10^{42}  \, T^5 \left[ 1 + \frac{T}{T_0}
\right]^{-3} \, h^{-2}_{2/3} \, \, \rm erg \, \, s^{-1} 
\en      

   \item[$\bullet$] {\bf $\mathbf{M_{ICM}-T}$ relation :}   

\be
M_{ICM}  =  2.405 \times 10^{12} \, T^3 \left[ 1 + \frac{T}{T_0} \right]^{-3/2} \,
h^{-2}_{2/3} \, \, \rm M_\odot
\en      

   \item[$\bullet$] {\bf $\mathbf{f_{gas}-T}$ relation :}

\be
f_{gas}  =  0.0792 \, T^{3/2} \left[ 1 + \frac{T}{T_0} \right]^{-3/2} \,
h^{-1}_{2/3}
\en      

   \item[$\bullet$] {\bf $\mathbf{y_0-T}$ relation :}

\be
y_0  =  3.506 \times 10^{-6} \, T^3 \left[ 1 + \frac{T}{T_0} \right]^{-3/2} \,
h^{-2/3}_{2/3} \:,
\en      
which can be translated in central temperature decrement as (using $T_{CMB} =
2.73 \, \rm K$ and the RJ approximation)
\be
\Delta T(0) =  -1.914\times 10^{-2} \, T^3 \left[ 1 + \frac{T}{T_0} \right]^{-3/2}
h^{-2/3}_{2/3} \, \rm mK \:.
\en

   \item[$\bullet$] {\bf $\mathbf{\Delta T(0)-L_X}$ relation :} \\

   \begin{description}

     \item[$\mathbf{\star}$] {\bf for $\mathbf{ T \gg T_0}$} (rich clusters
temperature range)~: \\
\be
L_X  =  2.206 \times 10^{45} \left(\frac{-\Delta T(0)}{1 \, \rm  mK}\right)^{4/3} \, h^{-10/9}_{2/3} \, \, \rm erg \, \, s^{-1} 
\en      
     \item[$\mathbf{\star}$] {\bf for $\mathbf{ T \ll T_0}$} (small groups temperature range)~: \\
\be
L_X  =  2.133 \times 10^{45} \left(\frac{-\Delta T(0)}{1 \, \rm  mK}\right)^{3/2} \, h^{-1}_{2/3} \, \, \rm erg \, \, s^{-1}. 
\en      

  \end{description}

\end{description}

%=========================================================================
\section{Infall velocity and mean temperature}
%=========================================================================
\label{app:b}

In this Appendix, we will find an expression for the infall velocity before the
shock (noted $v_1$) as a function of the mean temperature of the accreting
system. 
To this purpose, the following reasonning in two steps is
appropriate~: we first show that $v_1^2 \propto \phi_v$ provides a good
approximation and then derive a usefull expression for $\phi_v$.\\
We will then compare our result to hydrodynamic simulations, which will
validate the use of the analytic expression found in the groups mass range. 

\subsection{Reexpressing the infalling velocity}
%--------------------------------------------------

Let's assume that the gas inflow is stationary. In a spherical collapse
model, the gas is assumed to be at rest at a radius $r_{ta}$ (the so-called
``turn-around radius'') before falling into the cluster potential. Since the
gas will not be subjected to many processes changing its internal energy during
this inflow, we can assume that the flow is isentropic. Thus, applying
the Bernouilli equation between $r_{ta}$ and $r_v$ (just before the
shock) we obtain \citep[see \eg ][]{LL59}~: 
\be
\frac{v_1^2}{2} + \omega_1 + \phi_v = \omega_{ta} + \phi_{ta}
\label{eq:bern}
\en 
where $\omega$ denotes the gas specific enthalpy ($\omega = c_p T$
with $c_p$ the specific heat at constant pressure).
Denoting $\rho_{ta}$ and $\rho_1$
the densities at $r_{ta}$ and $r_v$ and , the following expression for $v_1$ can
easily be found~:  
\be
v_1^2 = -2 \phi_v \left[ 1 - \frac{\phi_{ta}}{\phi_v} + c_p
\frac{T_1}{\phi_v}\left( 1 -
\left(\frac{\rho_{ta}}{\rho_1}\right)^{2/3}\right)\right] .
\en
In the spherical collapse model, the density at $r_v$ is $\sim 8$ times
larger than at 
$r_{ta}$. 
% The typical overdensity in filaments being also much smaller
% than near the virial radius, the same simplification will occur with
% non-spherical inflow within large-scale filaments. 
We can thus write~:
\be
v_1^2 = -2 \phi_v \left[ 1 - \frac{\phi_{ta}}{\phi_v} + \frac{c_p}{4}
\frac{T_1}{\phi_v} \right] .
\en
Obviously, in large mass systems, the last ratio will be negligible {\it
i.e.\/,} the thermal energy of the gas will be negligible when compared to
its potential energy (the so-called ``cold inflow'' hypothesis). As gas is
preheated ({\it i.e.\/,} as $T_1$ rises) or 
as the mass of the system is lowered this assumption is questioned, since the
thermal content of the gas will be of the same order of the kinetic and
potential energy. Nevertheless, we will assume in the following that this
limiting behaviour only occurs at mass scales smaller than groups of
galaxies. We will show at the end of the appendix that hydrodynamic numerical
simulations of structure formation in different cosmological models validate
this last assumption.
%  If we assume a
% \emph{cold inflow}, the thermal energy content of the gas (represented by the
% term $w = c_p T$) can be neglected, which leads to     

We have thus~:
\be
v_1^2 =  -2 \phi_v \left( 1 - \frac{\phi_{ta}}{\phi_v} \right)\: .
\en
To take into account some uncertainties and shortcomings of this crude
treatment (as the fact that the infall is usually along filaments and so not
spherical), we will 
write the slightly more general formula~:
\be 
v_1^2  = - 2 \,\eta \, \phi_v,
\label{eq:v1}
\en
where $\eta$ is considered as a constant. $\eta$ could be 
calibrated against numerical simulations \citep[see {\it
e.g.\/,}][]{Miniati2000} or chosen so that the flow at infinite is the Hubble flow, as in
\citet{CMT98}. When used to compute the value of $T_0$ in
Sec.~\ref{sec:compt0}, $\eta$ is determined by the former method.

Note that the established fact that the square of the infall velocity is
proportionnal to the  
gravitational potential, both taken at the virial radius, is actually
not a surprise. Indeed, before the shock, the gas is thought to follow
the dark  matter evolution, its velocity being equal to the dark matter
one up to the virial radius \citep[see for example the figure 15
of][]{Frenketal99}. Since it has been shown that the dark matter infall velocity in a
spherical model scales as $M^{1/3}$, while the mean potential $\phi \propto M/R$
scales as $M^{2/3}$ (\citet{RK97}, based on the work of
\citet{Bertschinger85}), it is thus natural that $v_1^2 \propto M^{2/3} \propto \phi$.\\

\subsection{Linking $\phi_v$ with $T$}
%-------------------------------------

We now look for a useful expression for $\phi_v$ without assuming any
particular analytic expression for $\phi({\bf r})$. First, since the
Poisson equation states~:
\be
\nabla^2 \phi = 4 \pi \, G \, \rho,
\en
we can write any gravitational potential in the form~:
\be
\phi(x) = 4 \pi \, G \, \rho_c \, r_v^2 \, \tilde{\phi}(x),
\en
where $\rho_c$ is a characteristic density (the background cosmological
density for example), $r_v$ is the virial radius, $G$ the gravitational
constant and $\tilde{\phi}(x)$ is a dimensionless potential with $x = r/r_v$.
Therefore we have in particular,
\be
\label{eq:phiv}
\phi_v = 4 \pi \, G \, \rho_c \, r_v^2 \, \tilde{\phi}(1)\: .
\en
By using this as a general relation, we assumed implicitely
that $\tilde{\phi}(1)$ does not depend on the mass of the system.
However, numerical simulations indicate that the general
expression for the gravitational potential is indeed universal (\ie it takes
the same form in different cosmological models), but its normalisation
should depend slightly on the mass via the so-called concentration
parameter $c$ \citep[][hereafter NFW]{NFW97}. Indeed, using the so-called NFW
analytic expression, $\phi_v$ can be expressed  as \citep[see for
example][]{LM2001}\footnote{Note that the function $g(c)$ defined here
differs from the one adopted by \citet{LM2001}} :   
\be
\label{eq:virpotnfw}
\phi_v =  4 \pi \, G \, \rho_c \, r_v^2 \, g(c) \: .
\en
Since in these models, $c$ is a slowly varying function of the mass ($c(M) =
(M/M_{15})^{-0.086}$ for a $\Lambda$CDM universe at $z=0$, $M_{15}$ being
the mass in units of $10^{15} \, \rm M_{\odot}$) it is easily shown that
$g(c)\propto T^{0.012}$, that is to say, $g(c)$ can be considered as
constant to a very good approximation. As a consequence, ignoring the mass
dependency of $\tilde{\phi}(1)$ in equation~(\ref{eq:phiv}) is
reasonnable and we will thus use equation~(\ref{eq:phiv}) as a definition of the virial
radius gravitational potential.\\

An even more practical form for equation~(\ref{eq:phiv}) does involve the
(supposed isothermal) ICM temperature $T$. It is achievable through the use
of the virial theorem. It indeed provides us with general relations between
$r_v-M$ and $M-T$ in the following form~: 
\bea r_v & = & \left( \frac{3M}{4\, \pi \, \Delta \, \rho_c} \right)^{1/3}
\label{eq:rvm}\\  
M & = & M_* \left(\frac{T}{T_*} \right)^{3/2}\: , \label{eq:mt}
\ena 
where $\Delta$
is the virial radius overdensity (taken throughout this paper to be 
500, see Appendix~\ref{app:a}), $\rho_c$ is the critical density and $M_*$
and $T_*$ are constants defined in Appendix~\ref{app:a}.
% The value of $\xi$ is slightly controversial. Whereas theoretical
% considerations lead to $\xi=3/2$, some observations seem to show a
% steepening of the $M-T$ relation to $\xi \sim 1.8$ \citep{NMF2000}
% that could be due to early entropy injection. Since we want to discuss
% this very effect we will keep $\xi$ as an input, with \emph{only two
% possible values}, $3/2$ and $1.8$. 
As a conclusion, we can now write equation~(\ref{eq:phiv}) as~:
\bea
\phi_v & = &  4 \pi \, G \,\rho_c  \, \tilde{\phi}(1)\, \left(
\frac{3 \, M_*}{4 \pi \, \Delta \, \rho_c } \right)^{2/3}
\left(\frac{T}{T_*}\right) \\ 
& = & \phi_1 \left(\frac{T}{T_*}\right) \label{eq:phivt}
\ena

Eventually, by combining equations (\ref{eq:phivt}) and (\ref{eq:v1})
we reach the following relation describing the scaling of $v_1^2$ with $T$~:
\bea
v_1^2 & = & -2 \eta \phi_1 \left(\frac{T}{T_*}\right) \; .
\label{eq:v1t_app}
\ena

Equation~(\ref{eq:v1t_app}) is particularly interesting since
it can be directly compared to the  the results of hydrodynamic
numerical simulations. \citet{Miniati2000} have examined the
properties of shock waves around clusters and groups of galaxies in
two hydro+N-body realisations of two different cosmological
models. Having selected all structures with X-ray luminosity greater than $10^{41} \, \rm
erg/s$, they found that the relation between the mean temperature and
the infall velocity is given by an equation of the form (we only quote their
result for the $\Lambda$CDM cosmology, analogous expressions being found in
the SCDM cosmology)~:
\be
\label{eq:vsim}
v_s =  v_{s0} \left( \frac{T}{T_s} \right)^{0.52}.
\en  
with
\bea
v_{s0} & = & 1.9 \times 10^3 \, \rm km \, s^{-1} \, \, and \nonumber \\
T_s & = & 7.8\times10^7\,\rm K \nonumber
\ena
Their result spans a range from $10^6$ to several $10^7$ K, \ie from
poor groups to clusters, and is analogous to
equation~(\ref{eq:v1t_app})\footnote{We will take the exponent of
eq.~(\ref{eq:vsim}) to be 1/2.}.
 This, together with the fact that secondary infall
models predict the same type of relation as the one we deduced between $v_1$ and
$\phi_v$, validates the assumption of cold inflow that was made in deriving
equation~(\ref{eq:v1}), even for the group mass range.

%=============================================
\section{The shape factors $Q_X$, $Q_{SZ}$ and $Q_{M}$.}
%=============================================
\label{app:shapefactor} 

In this appendix, we evaluate the shape factors $Q_X$, $Q_{SZ}$ and $Q_M$,
which  enter the normalisation of the $L_X-T$, $y-T$ and $M_{ICM}-T$
relations. These quantities could in fact be considered as
renormalisation constants obtained by comparison with observations. We
will instead show here that they can be computed using a motivated model for
the adiabat profile. The model we present thus predicts not only the shape
but also the normalisation of the above correlations.

\subsection{Normalisation temperature}

The normalisations of the three relations cited above are defined in
equations~(\ref{eq:lxnorm}), (\ref{eq:micmpred}) and (\ref{eq:y0}), where the
temperature is in keV. The temperature at which the normalisation must be
evaluated is obtained by equating the temperature shape function in each of
these relations to 1. This is illustrated in figure~\ref{fig:normt}, which
shows the following functions (the last two ones being the same)~:
\bea
L_X(T) & = & \frac{T^5}{\left[ 1 + T/T_0\right]^3} \, \, \, \mathrm{(solid
\, \, line)} \nonumber \\
y(T) & = & \frac{T^3}{\left[ 1 + T/T_0\right]^{3/2}} \, \, \, \mathrm{(dashed
\, \, line)} \nonumber \\ 
M_{ICM}(T) & = & \frac{T^{3}}{\left[ 1 + T/T_0\right]^{3/2}} \, \, \,
\mathrm{(dashed \, \, line),} \nonumber 
\ena
and their intersection with the horizontal unit line. We can see that the
normalisation temperature of the three relations occurs at $T \sim 1.3
\, \rm keV$. These temperatures are well into the groups temperature range, and we
will thus compute the normalisation in this regime. First of all, we need an
motivated analytic adiabat profile to compute the integrals defining the
three normalisation constants.
 
%-----------------------------------------------------------
   \begin{figure} 
   \centering 
   \includegraphics[width=9cm]{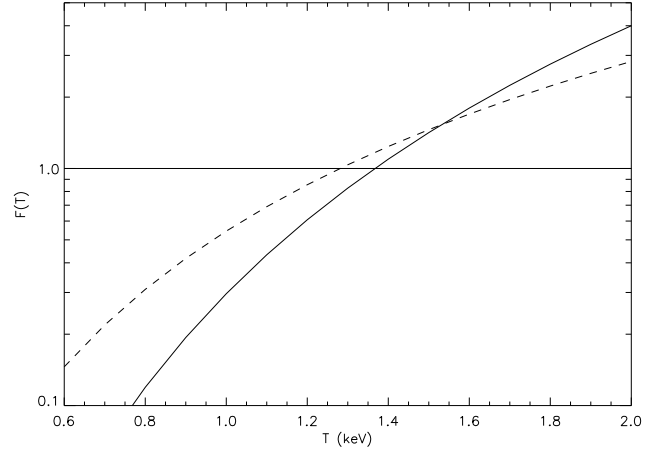}
   \caption{Normalisation temperature of the $L_X-T$ (solid line), $y-T$
    and $M_{ICM}-T$ (dashed line) relations. The three relations will thus be
   normalised at $T \sim 1.3 \, \rm keV$, in the groups mass range.
   \label{fig:normt}}
   \end{figure}
%-----------------------------------------------------------

\subsection{An analytic  model for the adiabat Profile}

Groups have a low surface brightness in X-rays and are thus difficult to
observe up to their virial radius. Obtaining a temperature profile at this
radius is obviously even more challenging. While the new generation of X-ray
satellites (Chandra and XMM-Newton in particular with its exceptional
sensitivity) is expected to produce constraints on low density regions,
ROSAT (whose energy band was perfectly adapted to the groups
observation) has provided data mostly on the inner parts of these systems
\citep[see {\it e.g.\/,}][]{HP2000}. Nevertheless, \citet{DJF96} have
published adiabat profiles for two cool clusters ($T \sim 1.5 \, \rm keV$)
nearly reaching the virial radius (as well as for even cooler groups, see
their figure 10), using a $\beta-$model for the density and the best-fit
power-law model for the temperature (despite large error bars). They found a
isentropic core, followed by a very modest increase compared to clusters
(where the virial value of the adiabat is at least 20 times higher than the
core value).

This can be easily understood, if one recalls that groups are thought to
accrete most of their gas adiabatically, and should thus have constant
adiabat profiles \citep{BBP99,TN2001}. However, this produces unacceptably
large temperature gradients in small groups and the quasi-absence of the
cooling-flow phenomenon, except in the largest clusters \citep[][see their
figures 4c and 7b]{TN2001}. It shows that somewhere between the center and
the virial radius, shocks must increase somewhat the adiabat, in order to
lower the temperature gradient, giving rise to an adiabatic inner core (much
larger than the core radius) followed by a modestly increasing adiabat
profile\footnote{Note that HCG62, one of the best examples of relaxed group
$(\overline{T}\sim 1 \, \rm keV$), has a steep declining temperature outside
the inner core \citep[see fig. 4 of][]{FP99}, but has also an obvious
cooling-flow.}.

One can seek an analytic expression for the adiabat profile by making use of
the isothermal $\beta-$model \citep{CFF76}. Suppose that the gas density
profile is describe by this model {\it i.e.\/,}
\be
n(x) = n_0 \left[1+\left(\frac{x}{x_0}\right)^2\right]^{-3\beta/2} \:,
\en
where the radii are rescaled in units of $r_v$ and $x_0$ is the core
radius. Assuming isothermality, the dimensionless adiabat profile is then 
\be
\label{eqapp:adiabgp}
f(x) = \left[1 + \left(\frac{x}{x_0}\right)^2\right]^\beta
\en

Using the semi-analytic model already discussed in the text  (Dos Santos 2001, in
preparation, see Sec.~\ref{sec:semianal}), we found that, in an external
preheated model (by QSOs or SNs), the adiabat profile of groups and clusters
can be described approximately by the last equation with $\beta = 1/2$ and a
core radius varying with mass. Even if the systems modelled are not
isothermal, this result is not surprising as the shocks are expected to leave a
linearly rising adiabat profile \citep[hence the value of $\beta$,
see][]{TC92,TN2001,BM2001}, while the isentropic accretion of gas leaves a
constant adiabat profile. Fixing $\beta$ to 1/2, the core radius $x_0$
determines the extend of the central isentropic core. We have found that $x_0
= 0.7$ reproduces approximately the adiabat profile of small groups in our
semi-analytic model, while $x=0.1$ reproduces its behaviour in rich
clusters. This is in line with the observational results of \citet{DJF96}
showing that the profiles are much shallower in groups than in clusters. Thus
we will use $\beta = 1/2$ and $x_0 = 0.7$ in eq.~(\ref{eqapp:adiabgp}) to
compute the values of the shape factors.

% Depending on the exact mass of the group and on the
% amount of external preheating, the parameters of eq.~(\ref{eqapp:adiabgp}) will
% change somewhat, but this won't change the derived values of the shape
% factors by more than 20-30 $\%$. We will thus use this expression for the
% adiabat profile to compute the shape factors in the following sections.

\subsection{The shape factor $Q_X$}
%----------------------------------
The theoretical shape factor is defined by:~
\bea
\label{eqapp:qxth}
Q_X  & = & \int_0^1 \frac{x^2 \, dx}{f^3(x)} \nonumber \\
& = & \int^1_0 x^2 \left[1+\left(\frac{x}{0.7}\right)^2\right]^{-3/2} \nonumber \\
& \sim & 0.1150  \nonumber
\ena
This value is very close to the best-fit value (0.12, considering $Q_X$ as an
adjustable parameter) and undistinguishable given the observational
errors. Although the normalisation has been computed in the groups mass
ranges, it recovers perfectly the slope and normalisation of rich clusters. We
will thus use $Q_X = 0.115$.

\subsection{The shape factor $Q_{SZ}$}
%-------------------------------------

The shape factor for the $y-T$ relation is given by :
\bea
\label{eqapp:qsz}
Q_{SZ} & = & \int_0^{\infty} \left[1 +\left(\frac{x}{0.7}\right)^2\right]^{-3/4} \, dx \\
& \sim & 1.835  \nonumber
\ena
Here again the normalisation is in very good agreement
with the observational data from \citet{Zhang2000}, despite the large
dispersion of the latter (see figure~\ref{fig:ytrelation}).
 
\subsection{The shape factor $Q_{M}$}
%------------------------------------

The shape factor for the $M_{ICM}-T$ relation is given by~:
\bea
\label{eqapp:micm}
Q_M & = & \int_0^{1}  x^2 \,
\left[1+\left(\frac{x}{0.7}\right)^2\right]^{-3/4} \nonumber \\ 
& \sim & 0.1915 \nonumber
\ena
The best-fit value to the data of \citet[][hereafter MME]{MME99} is $Q_M =
0.11$ {\it i.e.\/,} $\sim 40 \, \%$ lower than the value we compute. However,
apart from pure measurements uncertainties and known
biases\footnote{Note that testing their analysis method with hydrodynamic
numerical simulations, MME find that they are likely to overestimate the ICM
mass by a factor of $\sim 12 \, \%$, which increases the disagreement with
our $Q_M$ value by the same amount}, several systematic effects can cause
this disagreement : first, the major part of the 45 clusters they study are
not detected up to the radius $r_{500}$. Thus, they extrapolate the data
taking their best-fit slope out to $r_{500}$, which can be a source of
errors. Second, as MME explain, computing $r_{500}$ requires a model of the
potential or a rescaling to the value of a given cluster (they chose
A1795 for the scaling of their virial relations). If one of the assumptions
made is erroneous, it is possible that their numbers quote a radius $r_\Delta$
with $\Delta \neq 500$. Indeed, if $r_{500}$ is underestimated by only $20 \,
\%$, we obtain : 
\bea
Q_M & = & \int_0^{0.8} \left[1+\left(\frac{x}{0.7}\right)^2\right]^{-3/4} \nonumber \\
& \sim & 0.1136 \:,  \nonumber
\ena
which is very close to the best-fit value. We use this last value for $Q_M$
in the $M_{ICM}-T$ relation, as well as in the $f_{gas}-T$ relation.

The luminosity of a cluster is dominated by the core gas (since it scales
as the square of the density) and the SZ effect computed here is a central
value (taking into account the beam smearing effect), while the gas mass
depends mainly on the outer parts of the profile, where ourassumption of
self-similarity of the scaled adiabat profiles is more likely to break down
(see section~\ref{sec:ssassum}). This explains naturally why
the predictions for $Q_X$ and $Q_{SZ}$ are much more accurate than for
$Q_M$. Nevertheless, we consider the agreement between the theoretical
values and the observed normalisations very satisfying.
The fact that with the same model for the adiabat profile we can predict the
correct normalisations of the cluster correlations in two very different
wavebands is obviously a sign of coherence of the whole scheme. Since the
shape factors are predicted, \emph{the only parameter of the model is the
central entropy in groups $K_0$}, which was fit to the observations of
central adiabat value.

\end{document}